\documentclass[useAMS,usenatbib]{mnras}

\usepackage{graphicx}

\def\mnras{MNRAS}
\def\aap{A\&A}

\def\apj{APJ}
\def\apjs{APJS}
\def\apjl{APJL}
\def\aj{AJ}
\def\pasp{PASP}

\def\nat{Nature}
\def\araa{ARA\&A}
\def\actaa{Acta Astron}
\def\apss{Ap\&SS}

\newcommand{\degree}{\mbox{$^\circ$}}

\begin{document}

\title[LSQ14efd]{LSQ14efd: observations of the cooling of a shock break-out event in a type Ic Supernova.}
\author[C. Barbarino et al. ]{C. Barbarino$^{1,2}$, M.T. Botticella$^{2}$, M. Dall'Ora$^{2}$, M. Della Valle$^{2,3}$, S. Benetti$^{4}$, 
\newauthor J. D. Lyman$^{5}$, S. J. Smartt$^{6}$, I. Arcavi $^{7,8}$, C. Baltay$^{9}$, D. Bersier$^{10}$, M. Dennefeld$^{11}$,
\newauthor N. Ellman$^{9}$, M. Fraser$^{12}$, A. Gal-Yam$^{13}$, G. Hosseinzadeh$^{7,14}$, D. A. Howell$^{7,14}$,
\newauthor C. Inserra$^{6}$, E. Kankare$^{6}$, G. Leloudas$^{13,15}$, K. Maguire$^{6}$, C. McCully$^{7,14}$, A. Mitra$^{16,17}$,
\newauthor R. McKinnon$^{9}$, F. Olivares E.$^{18,19}$, G. Pignata$^{18,19}$, D. Rabinowitz$^{9}$, S. Rostami$^{9}$,
\newauthor K. W. Smith$^{6}$, M. Sullivan$^{20}$, S. Valenti$^{21}$, O. Yaron$^{13}$ and D. Young$^{6}$.
\\
$^{1}$Dip. di Fisica and ICRA, Sapienza Universit\'{a} di Roma, Piazzale Aldo Moro 5, I-00185 Rome, Italy\\
$^{2}$INAF- Osservatorio Astronomico di Capodimonte, Salita Moiariello 16, 80131 Napoli, Italy \\
$^{3}$ICRANet-Pescara, Piazza della Repubblica 10, I-65122 Pescara, Italy \\
$^{4}$INAF- Osservatorio Astronomico di Padova, Vicolo dell'Osservatorio 5, 35122 Padova, Italy \\
$^{5}$Department of Physics, University of Warwick, Coventry CV4 7AL, UK \\
$^{6}$Astrophysics Research Centre, School of Mathematics and Physics, Queen’s University Belfast, Belfast BT7 1NN, UK \\
$^{7}$Las Cumbres Observatory Global Telescope, 6740 Cortona Dr, Suite 102, Goleta, CA 93111, USA \\
$^{8}$Kavli Institute for Theoretical Physics, University of California, Santa Barbara, CA 93106, USA \\
$^{9}$Physics Department, Yale University, New Haven, CT 06511 \\
$^{10}$Astrophysics Research Institute, Liverpool John Moores University, Liverpool \\
$^{11}$Institut d’Astrophysique de Paris, 98bis Boulevard Arago, F-75014 Paris \\
$^{12}$Institute of Astronomy, University of Cambridge, Madingley Rd, Cambridge, CB3 0HA \\
$^{13}$Department of Particle Physics and Astrophysics, The Weizmann Institute of Science, Rehovot, 76100 Israel \\
$^{14}$Department of Physics, University of California, Santa Barbara, Broida Hall, Mail Code 9530, Santa Barbara, CA 93106-9530, USA \\
$^{15}$Dark Cosmology Centre, Niels Bohr Institute, University of Copenhagen, Juliane Maries vej 30, 2100 Copenhagen, Denmark \\
$^{16}$Sorbonne Universités, UPMC, Paris VI,UMR 7585, LPNHE, F-75005, Paris, France \\
$^{17}$CNRS, UMR 7585, Laboratoire de Physique Nucleaire et des Hautes Energies, 4 place Jussieu, 75005 Paris, France \\
$^{18}$Departamento de Ciencias Fisicas, Universidad Andres Bello, Avda. Republica 252, Santiago, Chile \\
$^{19}$Millennium Institute of Astrophysics, Santiago, Chile \\
$^{20}$School of Physics and Astronomy, University of Southampton, Southampton, SO17 1BJ, UK \\
$^{21}$Department of Physics, University of California, Davis, CA 95616, USA \\
}

\maketitle

\begin{abstract}

We present the photometric and spectroscopic evolution of the type Ic supernova  LSQ14efd, discovered  by the La Silla QUEST survey and followed by PESSTO. LSQ14efd was discovered few days after explosion and the observations cover up to $\sim$ 100 days.
The early photometric points show the signature of the cooling of the shock break-out event experienced by the  progenitor at the time of the supernova explosion, one of the first for a type Ic supernova. 
A comparison with type Ic supernova spectra shows that  LSQ14efd is quite similar to the type Ic SN 2004aw. These two 
supernovae have kinetic energies that are 
intermediate  between standard Ic explosions and those which are the most energetic explosions known (e.g. SN 1998bw). 
We  computed an analytical model for the light-curve peak and estimated the mass of the ejecta $6.3 \pm 0.5 \: M_{\odot}$, a synthesized nickel mass of $0.25 \: M_{\odot}$ and a kinetic energy of $E_{kin} = 5.6 \pm 0.5 \times 10^{51} \: \mathrm{erg}$. No connection between  LSQ14efd and a GRB event could be established.  
However we point out that the supernova shows some spectroscopic similarities with the peculiar SN-Ia 1999ac and  the SN-Iax SN 2008A. A core-collapse origin is most probable considering the spectroscopic, photometric evolution and the detection of the cooling of the shock break-out.
\end{abstract}  


\section{Introduction}\label{Introduction}

Supernovae (SNe) without hydrogen lines in their spectra are classified into two main types: SNe-Ia and SNe-Ib/c  (see \citealt{Filippenko1997} for a review). SNe-Ia originate from thermonuclear explosions of a carbon-oxygen white dwarf (CO-WD) reaching the Chandrasekhar mass. Two main channels for the origin of type Ia SNe have been proposed: a single degenerate scenario where the CO-WD in a binary system accretes matter from a companion star (\citealt{Wheeler1971}; \citealt{Whelan1973}), and a double degenerate scenario for which the SN is the result of a merging of two close WDs after orbital shrinking (\citealt{Tutukov1979}; \citealt{Iben1984}; \citealt{Webbink1984}). A third channel has been recently proposed by \citet{Katz2012} for which, in a WD triple system, the WDs approach each other and the collision is likely to detonate the WDs leading to a type Ia SN.
However, the detailed physics of the explosion are poorly understood and several models have been presented, from the supersonic detonation to the subsonic deflagration \citep{Hillebrandt2000}.
In the last years, several peculiar SNe-Ia were discovered (e.g. \citealt{Li2001}, \citealt{Valenti2014}), suggesting  the existence of a variety of explosion mechanisms and/or progenitor systems (e.g. \citealt{Mannucci2006}). Very bright objects,  with a luminosity $\sim 40\%$ brighter than  normal SN-Ia have been observed and are considered to be “super-Chandrasekhar” explosions (\citealt{Howell2006}; \citealt{Scalzo2010}; \citealt{Silverman2011}; \citealt{Taubenberger2011}). At the other extreme very faint events show unusual observational signatures (e.g., \citealt{Turatto1996};  \citealt{Foley2009}; \citealt{Perets2010}; \citealt{Kasliwal2010}; \citealt{Sullivan2011}).  A new explosion model was proposed for a particular sub-class of these sub-luminous SNe events that exhibit similarities to SN 2002cx \citep{Li2003}, called type Iax SNe (e.g 2005hk \citealt{Phillips2007}; 2008A \citealt{Foley2013}), which originate from the deflagration of a CO-WD that accretes matter from a companion He star.

Type Ib/c SNe originate from the gravitational collapse of a massive star for which the iron core cannot be supported by any further nuclear fusion reaction, or by electron degenerate pressure, hence collapsing into a neutron star or a black hole. They can be divided into two classes: SN-Ib which show He lines in their spectra and SN-Ic which do not.
Two main scenarios are considered for the progenitors of type Ib/c SNe (see reviews by \citet{Woosley2006} and \citealt{Smartt2009}): a single massive Wolf-Rayet (WR) star which has lost its hydrogen envelope, before the collapse of the core, through stellar winds or a binary system (see \citealt{Panagia1991} for an early suggestion) where the progenitor star loses its H (and He, in the case of SN-Ic) envelope through tidal stripping from the companion star. The measured masses of the ejecta seem to favour the majority being relatively lower mass binary stars, rather than very massive single WR stars (\citealt{Eldridge2013}, \citealt{Lyman2016}, \citealt{Cano2014}) and 
the data for iPTF13bvn seems to be more consistent with a binary 
system \citep{Bersten2014,Fremling2014,Eldridge2015}. 

Unlike SNe-IIP and IIb, the progenitors have not been commonly identified in pre-explosion images \citep[see][]{Eldridge2013,Smartt2015b}. One probable detection exists for the progenitor of a Ib SN : namely iPTF13bvn by \cite{Cao2013}, which has been studied further by 
\cite{Groh2013}, \cite{Fremling2014}, \cite{Bersten2014} and \cite{Eldridge2015}.

Almost two decades of observations have allowed us to divide the SN-Ic population into two sub-classes: \textit{standard} SNe-Ic, characterized by kinetic energies of $E_{k} \sim 10^{51}$ erg and broad-line (BL),  with  ejecta velocities  of order $\sim 0.1c$ which therefore implies significantly higher kinetic energy \cite[$E_{k} \sim  10^{52}$ erg, e.g.][]{Nakamura2001}. Some high energy, SNe-Ic-BL have been convincingly linked to gamma ray bursts (GRB; see \citealt{Kovacevic2014} for an updated census of GRB-SNe), while the majority of SNe-Ic are not associated with GRBs (e.g. the ratio GRB/SNIbc is $<3\%$, \citealt{guetta07}).
Some SNe-Ic, such  as SN 2004aw \citep{Taubenberger2006} and SN 2003jd \citep{Valenti2008a} show physical properties in between those of standard Ic events and SNe-Ic-BL, therefore suggesting the existence of a wide diversity in SNe-Ic in terms of expansion velocity of the ejecta, peak  luminosity, and kinetic energy (\citealt{Elmhamdi2006,Modjaz2015}).
In this scenario, it is not clear if the broad variety of observed SNe-Ic and SNe-Ic-BL is due to different sub-classes of SNe-Ic originating from different classes of progenitors, or if is representative of an existing continuum of properties among the different SN-Ic types (\citealt{MDV2011}; \citealt{Modjaz2014} \citealt{Prentice2016}).

The large fraction of peculiar objects, for both SN-Ia and Ic classes, which are now being 
found has led to cases of ambiguity in determining the physical
origin of these hydrogen and helium poor objects. In many cases
the physical origin in a thermonuclear or core-collapse explosion is debated  e.g. SN 2002bj (\citealt{,Poznanski2010}); SN 2004cs (\citealt{Rajala2004,Rajala2005,Leaman2011}) and SN 2006P (\citealt{Serduke2006,Leaman2011,Li2011}).
Usually the photometric analysis does not solve the ambiguity that can arise from the early spectroscopy,  for example \citet{Cappellaro2015} found a  $\sim 40\%$  difference in the the classifications of SN-Ib/c classification using PSNID within the
SUDARE survey. 

 Nebular spectra may help, 
 in revealing nucleosynthetic products in the interior part of the star, which are very different in 
 the two mechanisms but it is possible to observe them only for the brigthest and nearest sources. 

In some cases,  early photometric observations may be able to 
detect the cooling of the shock break-outs which can constrain the progenitor system.
One of the main signatures of core-collapse SNe (CC-SNe) is represented by the early emission of X-ray and/or ultraviolet radiation which traces
 the break-out of the SN shock-wave through the stellar photosphere. After the envelope of the star has been shock heated it starts cooling and it creates an early peak in the optical passband. This event represents the initial stages of a SN event and it typical shows very short duration, from minutes to hours, which makes the detection difficult and has resulted in the number of observed cooling of the shock break-outs being few in number (e.g. SN 1987A, \citealt{Arnett1989}; SN 1993J, \citealt{Lewis1994}; SN 2006aj, \citealt{Campana2006}; SN 2008D, \citealt{Mazzali2008}; \citealt{Soderberg2008} and SN 2011dh \citealt{Arcavi2011}).
 This early emission can be interpreted also as an extended envelope or due to outwardly mixed $^{56}Ni$ as investigated for the peculiar type Ib/c SN 2013ge \citep{Drout2016}.

The cooling of the shock break-out in a type Ia SN explosion in a WD is too dim and fast to be detectable for extragalactic events (\citealt{Nakar2012}; \citealt{Rabinak2012}) but an 
early UV excess in type Ia SNe is predicted for certain  binary progenitor systems.
\citet{Kasen2010} shows that the collision between the SN ejecta and its companion star should produce detectable radiative diffusion from deeper layers of shock-heated ejecta causing a longer lasting optical/UV emission, which exceeds the radioactively powered luminosity of the SN for the first few days after the explosion. \\
Here present an extensive data set for LSQ14efd which was discovered
by the La Silla QUEST survey (LSQ; \citealt{Baltay2013}),  and monitored by the  Public ESO Spectroscopic Survey of Transient Object (PESSTO; \citealt{Smartt2015})\footnote{www.pessto.org}.
Our analysis suggests that LSQ14efd is most similar to SN 2004aw which was originally classified as type Ia SN and re-classified as a type Ic after a long-term followup.
We show a detection of the cooling envelope emission after the shock break-out. The photometric and spectroscopic analysis of LSQ14efd shows this is a SN-Ic.
However, LSQ14efd shows some ambiguities in its spectroscopic classification with
similarities to SNe Iax such as SN 2008A. 
LSQ14efd is also interesting object because it shows intermediate properties between ``standard'' and very energetic SNe-Ic events.

This paper is organized as follows: in Section 2 we present the discovery and the classification of  LSQ14efd and we discuss the properties of the host galaxy, the distance and the extinction; in Section 3 we present the optical photometric evolution of  LSQ14efd and compare its colour evolution and bolometric light curve with those of other Type Ic and Ia SNe. In Section 4 we present the optical spectroscopic observations and comparison with other SNe. In Section 5 we model the light curve peak to estimate the main physical parameters of the explosion such as ejected mass, kinetic energy and nickel mass. In Section 6 we summarize our discussion and present our conclusions.

\section{Discovery and host galaxy}\label{discovery}

\begin{figure}
  \centering
  \includegraphics[scale=0.5]{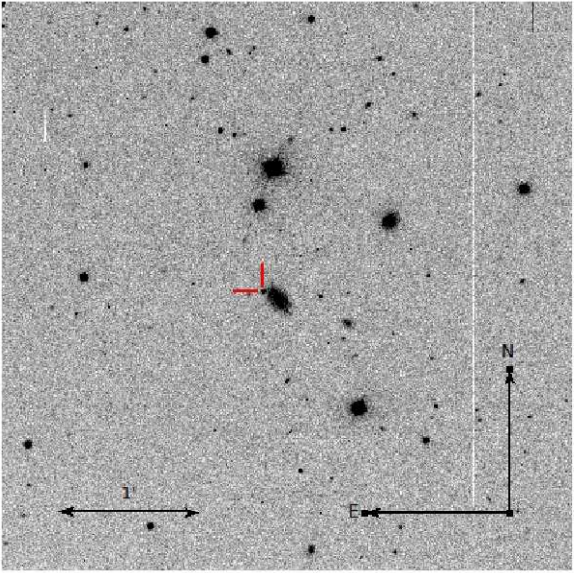}
  \caption{An image of LSQ14efd and the anonymous host galaxy, acquired with the New Technology Telescope and the EFOSC2 camera. The field of view is $4 \times 4$ arcmin$^2$.}
 \label{FC}
\end{figure}

LSQ14efd was discovered in an anonymous galaxy on 2014 August 17 UT (MJD=56886.81) at the coordinates R.A.$=03$h$:35$m$:38.74$s and DEC.$=-58\degree:52':38.3"$. After the discovery all images from LSQ archive were checked and we found the SN was already visible on the image of August 13, but not in images from August 09, with both images having the same depth. We thus consider the explosion date as August 11, with an uncertainty of 2 day. On 2014 August 18, \citet{Tartaglia2014a} classified LSQ14efd as a Type II SN with an unknown phase, as part of PESSTO. Due to the poor signal to noise (S/N) of the first spectrum, new observations were performed the day after and LSQ14efd was re-classified as a Type I SN around maximum \citep{Tartaglia2014b}. They also reported that the best match of the spectrum of  this transient was obtained  with the peculiar Type Ic SN 2003jd. LSQ14efd is located in an outer region of the host galaxy (see Fig \ref{FC}).
The distance to the galaxy is not available in the literature. The 2D spectroscopic frames shows the presence of the host galaxy spectrum for which it was possible to identify the H$\alpha$ emission, at a redshift of $z = 0.0672 \pm 0.0001$.  Assuming $H_{0}= 70 \: \rm km \: s^{-1} \: Mpc^{-1}$, we then calculate a distance modulus of $\mu= 37.35 \pm 0.03$ mag. The Galactic reddening towards LSQ14efd is estimated from   \citet{Schlegel1998} to be $E(B-V)= 0.0376 \pm 0.0015$ mag.  We considered the internal reddening of the host galaxy as negligible since there is no clear evidence of NaID absorption in the spectra nor significant
reddening of the spectrum continuum. 
Assuming a \citet{Cardelli1989} reddening law ($R_{V}=3.1$)  we estimate the total V-band extinction towards LSQ14efd to be $A_{V}= 0.12$ mag.

\subsection{Search for an associated GRB}

We have investigated the possibility that  LSQ14efd could be related with a GRB event. We have examined the Fermi Gamma-ray Burst Monitor\footnote{http://gammaray.msfc.nasa.gov/gbm/} (GBM) and the SWIFT\footnote{http://swift.gsfc.nasa.gov/archive/} catalogs in a period of time $< 20$ days from the occurence of the SN event. The time interval was chosen since it corresponds to the threshold of $\ge 95 \%$ confidence level for the association between GRB and type Ib/c SNe \citep{Kovacevic2014}.
In this range of time, we found $3$ detections, but none of them were spatially coincident, within the errorbox of the GRB detection, with the SN position. Hence, no association can be found between LSQ14efd and a GRB event.

\section{Photometric evolution}\label{Photsec}

\subsection{Data sample and reduction}

A photometric monitoring campaign for LSQ14efd, at optical wavelengths, was conducted over a period of $~100$ days post-discovery, covering $40$ epochs, using multiple observing facilities.

The $B \: V \: R \: I$ Johnson-Cousins data were collected with: the $1$m from Las Cumbres Observatory Global Telescope Network (LCOGT, Siding Spring, Australia) equipped with the SBIG Camera ($BV$, $10$ epochs); the $1$m from LCOGT (South African Astronomical Observatory, South Africa) equipped with the SBIG Camera ($BV$, $7$ epochs); the $1$m from LCOGT (Cerro Tololo, Chile) equipped with the Sinistro Camera ($BV$, $3$ epochs); the $3.58$m ESO New Technology Telescope (NTT, La Silla, Chile) equipped with the EFOSC2 (ESO Faint Object and Spectrograph Camera) camera ($BVRI$, $3$ epochs); the $1$m Schmidt telescope (La Silla, Chile) equipped with the QUEST camera ($V$, $10$ epochs) $0.6$m ESO TRAnsiting Planets and PlanetesImals Small Telescope (TRAPPIST, La Silla, Chile), equipped with TRAPPISTCAM ($VRI$, $6$ epochs).

$g'r'i'$ images were collected with: the $1$m from LCOGT (Siding Spring, Australia) equipped with the SBIG Camera ($BV$, $10$ epochs); the $1$m from LCOGT (South African Astronomical Observatory, South Africa) equipped with the SBIG Camera ($BV$, $7$ epochs) and the $1$m from LCOGT (Cerro Tololo, Chile) equipped with the Sinistro Camera ($BV$, $3$ epochs). A summary of the telescopes and the intruments characteristics are presented in Table \ref{phtel}.

\begin{table*}
\caption{Summary of the characteristics of the instruments used during the photometric follow up.\label{phtel}}
\begin{footnotesize}
\begin{tabular}{lclllcll}
\hline
Telescope  & Camera & Pixel scale & Field of view & Filters$^{a}$ & \# of epochs \\
 &  & [arcsec/pix] & [arcmin] &   \\
\hline
NTT (3.58m) & EFOSC2  & 0.24  & 4 $\times$ 4  & $B, V, R$; $u, g,r, i$          & 12 \\
COJ (1m) & SBIG	& 0.46 & 15.8 & $B, V, g, r, i$ & 10 \\
CPT (1m) & SBIG & 0.46 & 15.8 & $B, V, g, r, i$ & 7 \\
LSC (1m) & Sinistro & 0.38 & 26.5 $\times$ 26.5 & $B, V, g, r, i$ & 3 \\
ESO (1m) & QUEST & 0.87 & $2160 \times 2760$ & $V$ & 15 \\
TRAPPIST (0.60m) & TRAPPISTCAM     & 0.65    & 27 $\times$ 27   & $B$, $V$, $R$              &  4 \\
\hline
\end{tabular}
\\[1.5ex]
NTT = New Technology Telescope with the optical camera ESO Faint Object Spectrograph EFOSC2; COJ = $1$m Las Cumbres Observatory Global Telescope (LCOGT) with the SBIG camera, site in Siding Spring Observatory; CPT = $1$m LCOGT with the SBIG camera, site in South African Astronomical Observatory; LSC = $1$m LCOGT with the Sinistro Camera, site in Cerro Tololo; ESO = $1$m Schmidt telescope, site in La Silla Observatory; TRAPPIST = TRAnsit Planets and PlanetesImals Small Telescope. \\
$^{a}$ The NTT i filter is Gunn.

\end{footnotesize}
\end{table*}

Data pre-reduction followed the standard procedures of bias, overscan, flat-field corrections and trimming in the \texttt{IRAF} \footnote{IRAF is distributed by the National Optical Astronomical Observatory, which is operated by the Association of Universities for Research in Astronomy, Inc., under cooperative agreement with the National Science Foundation.} environment.
Johnson $B \: V$ and $g'r'i'$ calibrated magnitudes of $45$ reference stars were obtained through the AAVSO Photometric All-Sky Survey (APASS) \citep{Munari2014}. The internal accuracy of the APASS photometry, expressed as the error of the mean of data obtained and separately calibrated over a median of four distinct observing epochs and distributed between 2009 and 2013, is $0.013$, $0.012$, $0.012$, $0.014$, and $0.021$ mag for the $BVg'r'i'$ bands, respectively.
In our knowledge no other star catalogues were available for the field of the SN. Johnson-Cousins $RI$ photometry was estimated trasforming the $g'r'i'$ filters through the \cite{Lupton2005} transformation equations.

The \texttt{QUBA} pipeline \citep{Valenti2011} was used for most of the photometric measurements. This pipeline performs a point-spread-function (PSF) fitting photometry, based on \texttt{DAOPHOT} \citep{Stetson1987}, on both the SN and the sorted reference stars. \texttt{QUBA} allows one to model the background with a polynomial surface, to treat the cases in which the SN is embedded in a spatially varying background.
After some empirical tests, we found that a $4th$-order polynomial model gives good results for the background subtraction, considering the high S/N of the SN in the images.

Photometry on LSQ data, for the pre-discovery epochs, was performed with the stand-alone version of the \texttt{DAOPHOTIV/ALLSTAR} software, which allows 
cross-correlation of the measurements on the individual images, and produces a light curve with respect to a reference epoch. This approach is particularly useful when dealing with point sources with a low S/N ratio, as in the case of the first and last epochs of the LSQ data of LSQ14efd. The LSQ filter is a custom broadband filter centered on 5534 \AA, ranging from 4000 to 7000 \AA \citep{Baltay2012}. It is customary to transform LSQ instrumental magnitudes to Johnson $V$ magnitudes, by computing a ($V,B-V$) colour equation, where the coefficients are estimated on selected reference stars in the field, for which standard $B,V$ magnitudes are available. In our case, instrumental LSQ magnitudes were transformed to standard Johnson $V$ magnitudes taking advantage of the $B-V$ colour curve, estimated with the other telescopes. 

At all epochs, we estimated  $K-$correction factors from our spectroscopy and we found out that they are negligible, compared with the photometric uncertainty,
therefore none were applied. 

The full photometric measurements of the SN in the $BVRI$ and $g'r'i'$ are listed in Table \ref{phot}. The Johnson-Cousins $BVRI$ photometry is reported in Vega magnitudes, while the $g'r'i'$ photometry is reported in the AB magnitude system.

\begin{table*}
\caption{Optical photometry in Johnson-Cousins filters, in $g'r'i'$ bands, with associated errors in parentheses.\label{phot}}
\centering
\begin{tabular}{llccccccc}
\hline
Date & $\rm MJD$  & $B$ & $V$  & $R$  & $I$ & $g'$ & $r'$ & $i'$  \\
& & (mag) & (mag) & (mag) & (mag) & (mag) & (mag) & (mag)  \\
\hline
20140813 & 56882.48 & & 20.03 (0.09) & & & & & \\
20140815 & 56884.44 & & 20.15 (0.12) & & & & & \\ 
20140817 & 56886.48 & & 19.82 (0.04) & & & & & \\
20140819 & 56888.78 & 19.88 (0.13) & 19.43	(0.14) & & & 19.69 (0.14) & 19.47 (0.10) & 19.68	(0.19) \\
20140821 & 56890.47 & & 19.18 (0.06) & & & & & \\
20140822 & 56891.47 & & 19.17 (0.04) & & & & & \\
20140823 & 56892.76 & 19.43 (0.09) & 19.06	(0.15) & & & 19.33 (0.09) & 19.17 (0.08) & 19.25 (0.14) \\
20140826 & 56895.72 &	19.42 (0.11) & 19.04 (0.09) & 18.99	(0.17) & 18.83	(0.25) & & & \\
20140831 & 56900.74 &	19.46 (0.11) & 18.97 (0.15) & & & 19.11	(0.18) & 18.87 (0.12) & 19.02	(0.19) \\
20140903 & 56903.81 &	19.79 (0.15) & 18.96 (0.09) & 18.84	(0.19) & 18.66 (0.15) & & & \\
20140905 & 56906.12 &	20.22 (0.15) & 19.15 (0.07) & & & 19.51 (0.08) & & 19.09 (0.12) \\
20140907 & 56907.41 & & 19.16 (0.05) & & & & & \\
20140908 & 56908.96 & & 19.24 (0.18) & & & 19.83 (0.13) & 18.98 (0.07) & 19.11	(0.21) \\
20140911 & 56912.32 & & 19.39 (0.21) & 18.99 (0.28) & 18.81	(0.08) & & & \\
20140912 & 56912.42 & & 19.41 (0.05) & & & & & \\
20140913 & 56913.64 &	20.51 (0.21) & 19.48 (0.06) & & & 20.05	(0.11) & 19.12 (0.04) &	19.19	(0.06) \\
20140915 & 56916.73 & 20.71 (0.22) & 19.66	(0.20) & 19.14 (0.09) & 18.85 (0.12) & &  & \\
20140916 & 56916.41 & & 19.69 (0.07) & & & & & \\
20140917 & 56917.60 & 20.71 (0.21) & 19.74 (0.15) & & & 20.62 (0.21) & 19.29 (0.15) & 19.21 (0.21) \\
20140920 & 56920.96 &	& 19.89 (0.19) & & & 20.95 (0.28) & 19.50 (0.17) & 19.34 (0.16) \\
20140921 & 56922.32 & &	20.08 (0.21) & 19.66 (0.21) & 19.23 (0.19) & & & \\
20140922 & 56923.11 &	20.82 (0.21) & 20.09 (0.19) & & & 20.99 (0.21) & 19.56 (0.16) & 19.41 (0.22) \\
20140927 & 56927.60 & &	20.31 (0.21) & & & 21.16 (0.21) & 19.84 (0.21) & 19.74	(0.19) \\
20140928 & 56928.39 & & 20.14 (0.11) & & & & & \\
20140930 & 56930.40 & & 20.38 (0.08) & & & & & \\
20141002 & 56932.38 & & 20.39 (0.09) & & & & & \\
20141005 & 56936.03 & & 20.67 (0.21) & & & 21.57 (0.22) & 20.11 (0.23) & 19.94 (0.15) \\
20141009 & 56939.96 & & & & & & 20.41 (0.19) & \\
20141010 & 56940.52	& &	20.76 (0.26) & & & & 20.44 (0.26) &	20.01 (0.23) \\
20141011 & 56942.27 & &	20.98 (0.22) & 20.42 (0.22) & 20.11 (0.21) & & & \\
20141017 & 56948.06 & & 21.02 (0.23) & & & & &	20.28 (0.25) \\
20141018 & 56948.54 & & & & & 21.74	(0.21) &	20.66 (0.22) & 20.25 (0.23) \\
20141021 & 56952.23 & &	21.39 (0.21) & 20.86	(0.23) &	20.23 (0.21) & & & \\
20141022 & 56952.71 & & 21.26 (0.21) & & & & & \\
20141025 & 56955.70 & & & & & & 20.81 (0.23) & 20.42 (0.22) \\
20141029 & 56960.17 & &	21.41 (0.22) & & & &	21.02 (0.21) & \\
20141101 & 56963.24 & & 21.62 (0.21) & 20.9 (0.22) & 20.42 (0.21) & & &  \\
20141102 & 56964.22 & & & & & & & 20.87 (0.21) \\
20141111 & 56973.20 & & & 21.09 (0.12) & 20.55 (0.11) & & & \\
20141118 & 56979.15 & & & & & & 21.32 (0.22) & 21.32 (0.24) \\
\hline
\end{tabular}
\\[1.4ex]
\end{table*}

\subsection{Data analysis}\label{photoanalysis}

\subsubsection{Early evolution}

\begin{figure}
	\centering
	\includegraphics[scale=0.45]{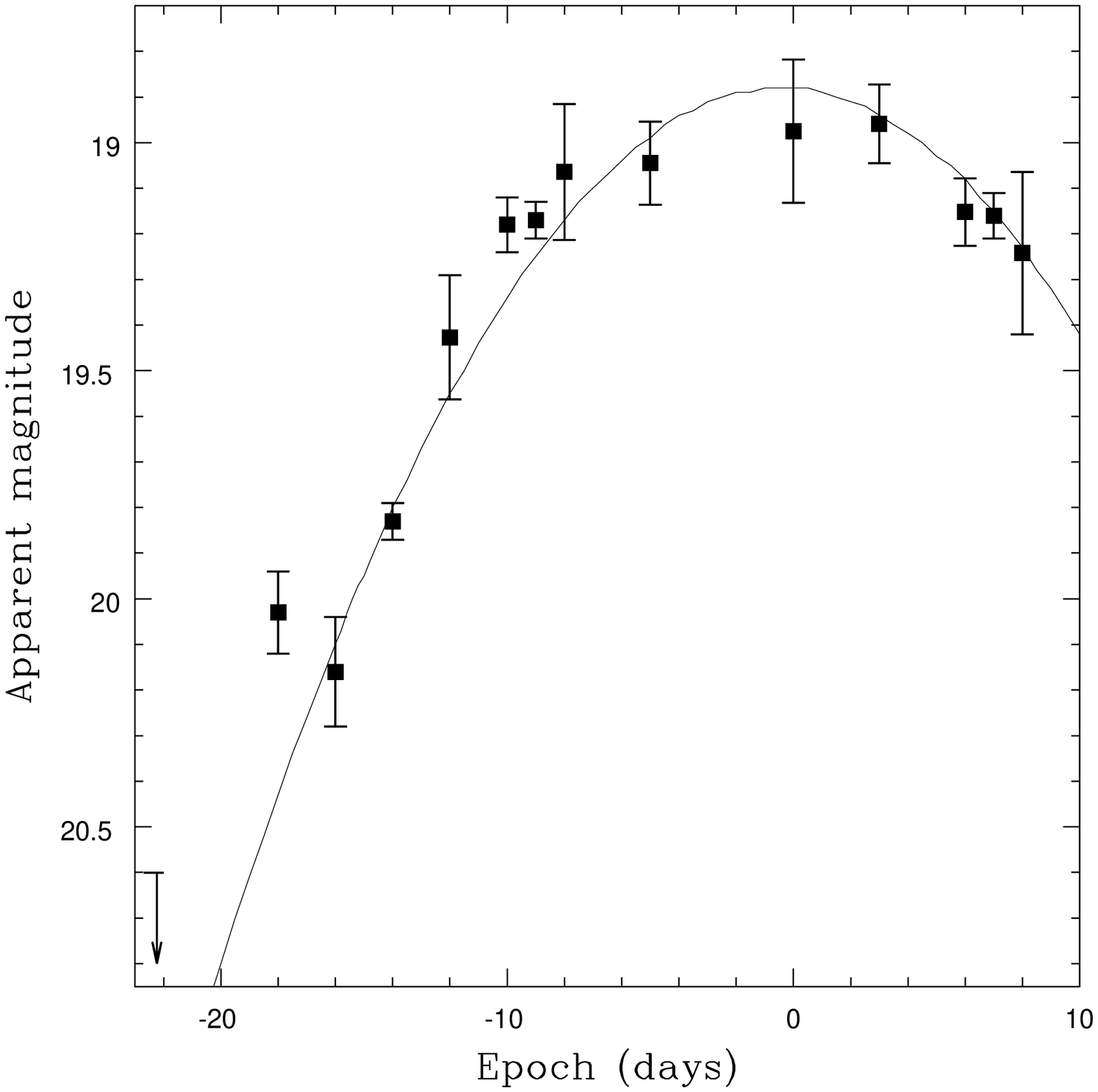}
	\caption{Fit of the light curve in the apparent V band magnitude of LSQ14efd. Epochs refer to the V-maximum. The pre-explosion limit is represented with the arrow.}
	\label{fit_V}
\end{figure}

The early points in the $V$ band show an initial decline before the rising of the peak. 
Fig. \ref{fit_V} shows a 2nd order polynomial fit of the early stage of the V band light curve without taking into account the first point. We note that it deviates significantly, 1$\sigma$, from the expected rise. A pre-explosion limit has been calculated from the image of August 09. The estimated limit is $20.6 \pm 0.2$ mag and it is represented by the arrow in Fig. \ref{fit_V}.

\begin{figure*}
  \centering
  \includegraphics[scale=0.36]{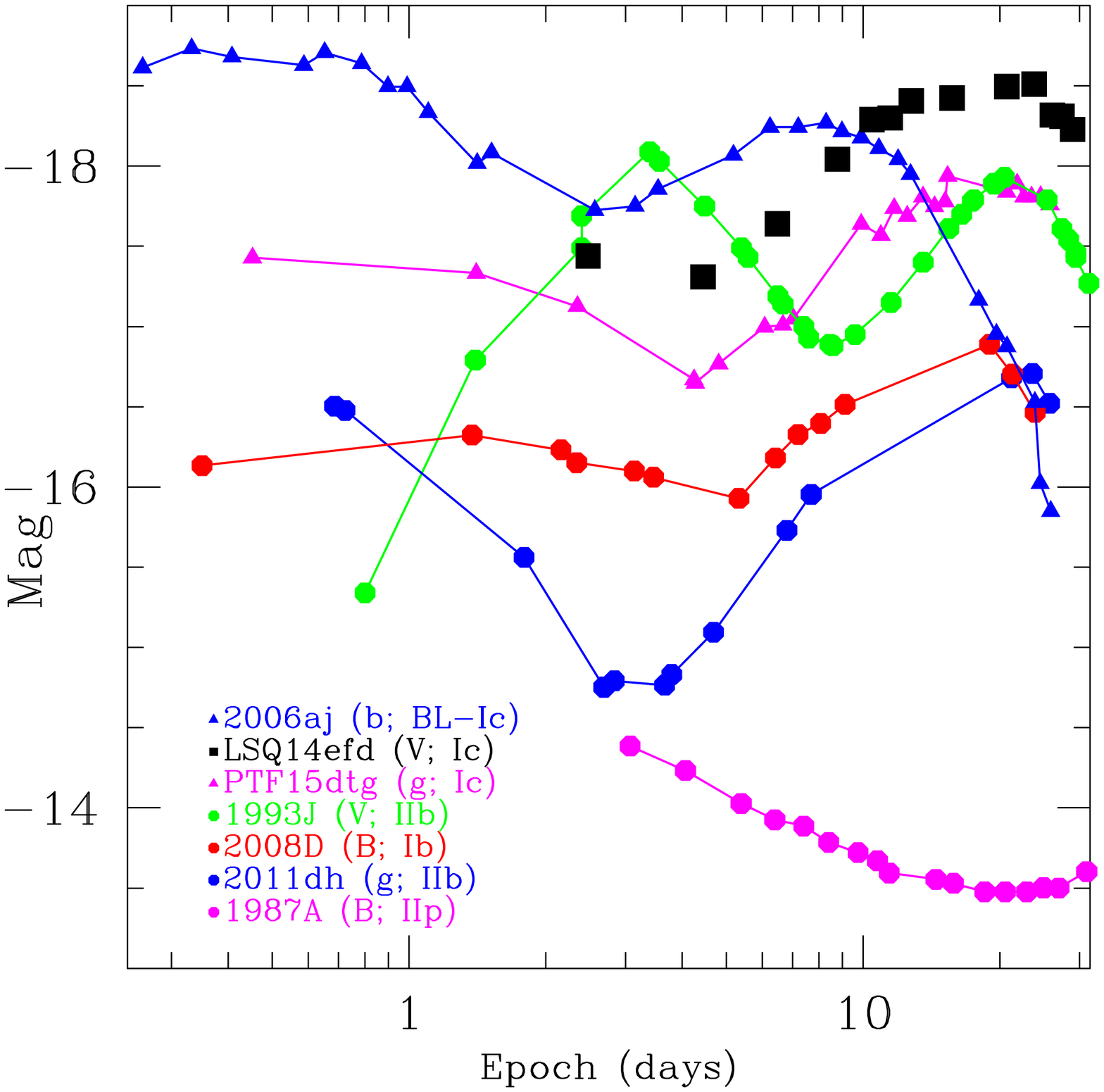}
  \includegraphics[scale=0.36]{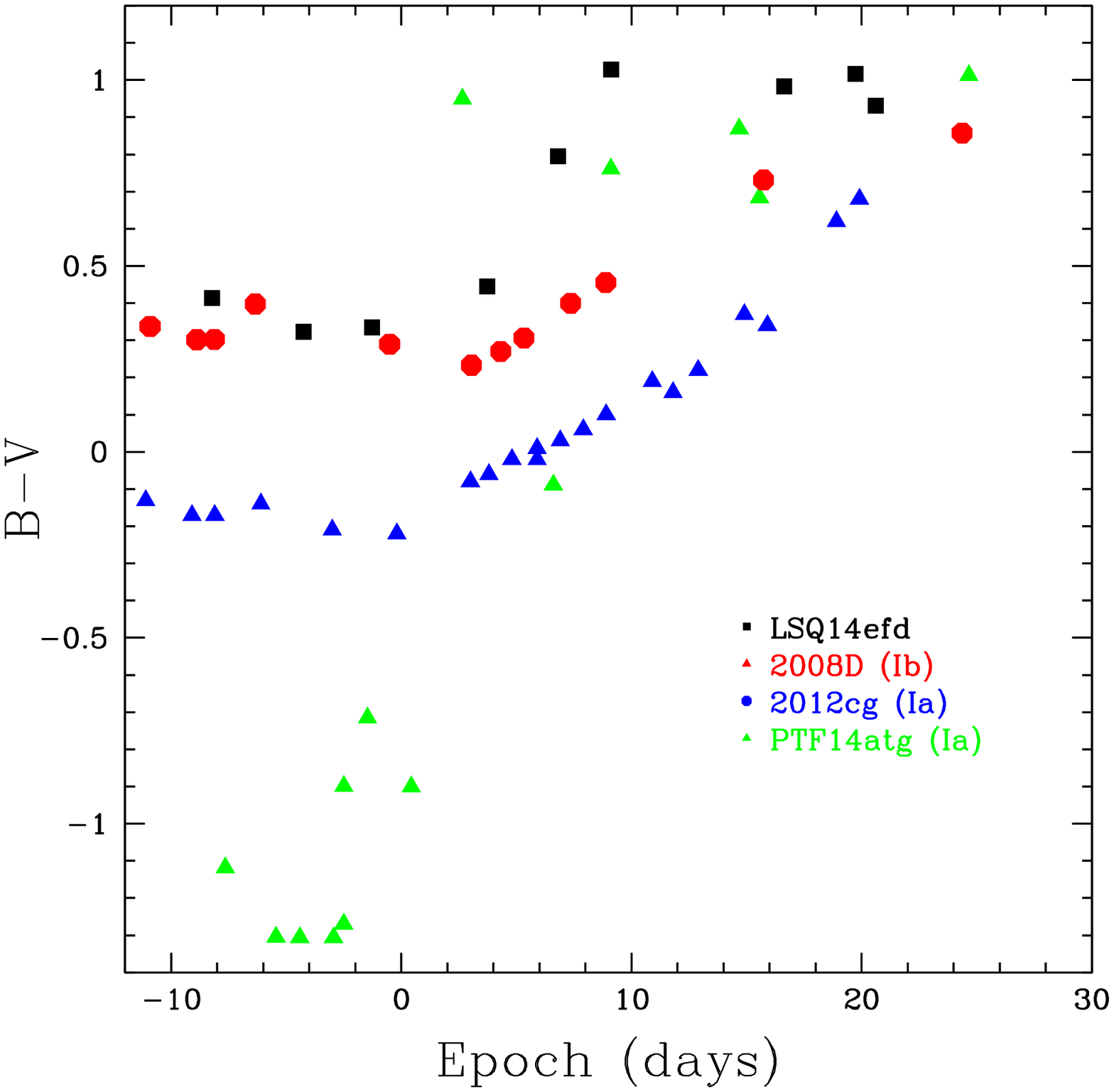}
  \caption{Left panel: qualitative comparison of the detection of the cooling of the shock break-out of LSQ14efd with that of some CC-SNe (SN 1993J, \citealt{Lewis1994}; SN 2008D, \citealt{Soderberg2008}; SN 2011dh, \citealt{Arcavi2011}; SN 2006aj \citealt{Campana2006}; iPTF15dtg, \citealt{Taddia2016} and SN 1987A \citealt{Shelton1993}). The compared bands are in the label. Phases refer to the days since explosion, in logarithmic scale. Right panel: comparison of the $B-V$ colour evolution of LSQ14efd with type Ia SNe iPTF14atg \citep{Cao2015} and 2012cg \citep{Marion2015} and type Ib SN 2008D \citep{Soderberg2008}. Phases refer to the days from the B-maximum.}
  \label{shock}
\end{figure*}

We interpret this decline as  the cooling that occurs soon after the shock break-out event, similar to that 
observed in other CC-SNe (e.g. SN 1993J, \citealt{Lewis1994}; SN 2006aj \citealt{Campana2006}; SN 2008D, \citealt{Soderberg2008}; SN 2011dh, \citealt{Arcavi2011}; iPTF15dtg \citealt{Taddia2016}), as shown in Fig. \ref{shock}.
A direct comparison of the decline observed in LSQ14efd with CC SNe shows good agreement with the behaviour observed in  previous cases where very
early photometry exists, in particular there is good relative agreement 
with SN 2008D and similar also to SNe 1993J and 2006aj (see, left panel of Fig. \ref{shock}).

An early UV emission pulse has been predicted also for type Ia SNe which generate from an interactive binary system (\citealt{Pakmor2008}, \citealt{Kasen2010}) and it has been observed recently (e.g. SN 2012cg, \citealt{Silverman2012}; \citealt{Marion2015} and iPTF14atg \citealt{Cao2015}).

We do not have UV data to directly compare the UV excess so we investigated the $B-v$ colour evolution.
The $B-V$ colour evolution is compared in Fig \ref{shock}
 to further investigate the early phase UV excess.
We notice that the $B-V$ colour evolution of iPTF14atg shows a pre- maximum value  around $-1.5$ and $-1$ around 10 days and it then sharply increases above $0$ around maximum. SN 2012cg shows a different $B-V$ colour evolution, being constant around $0$ until maximum
and  smoothly reaching a value $\sim 1$ at 20 days.
The  LSQ14efd $B-V$ evolution shows a different behaviour with respect to iPTF14atg, while shows a qualitative similar trend with respect to SN 2012cg but redder, since it differs of $\sim 0.5$ at all epochs.
Instead, the $B-V$ colour evolution of SN 2008D in pre- maximum phases is similar to that of LSQ14efd. 

\subsubsection{Late evolution}

\begin{figure*}
  \centering
  \includegraphics[scale=0.35]{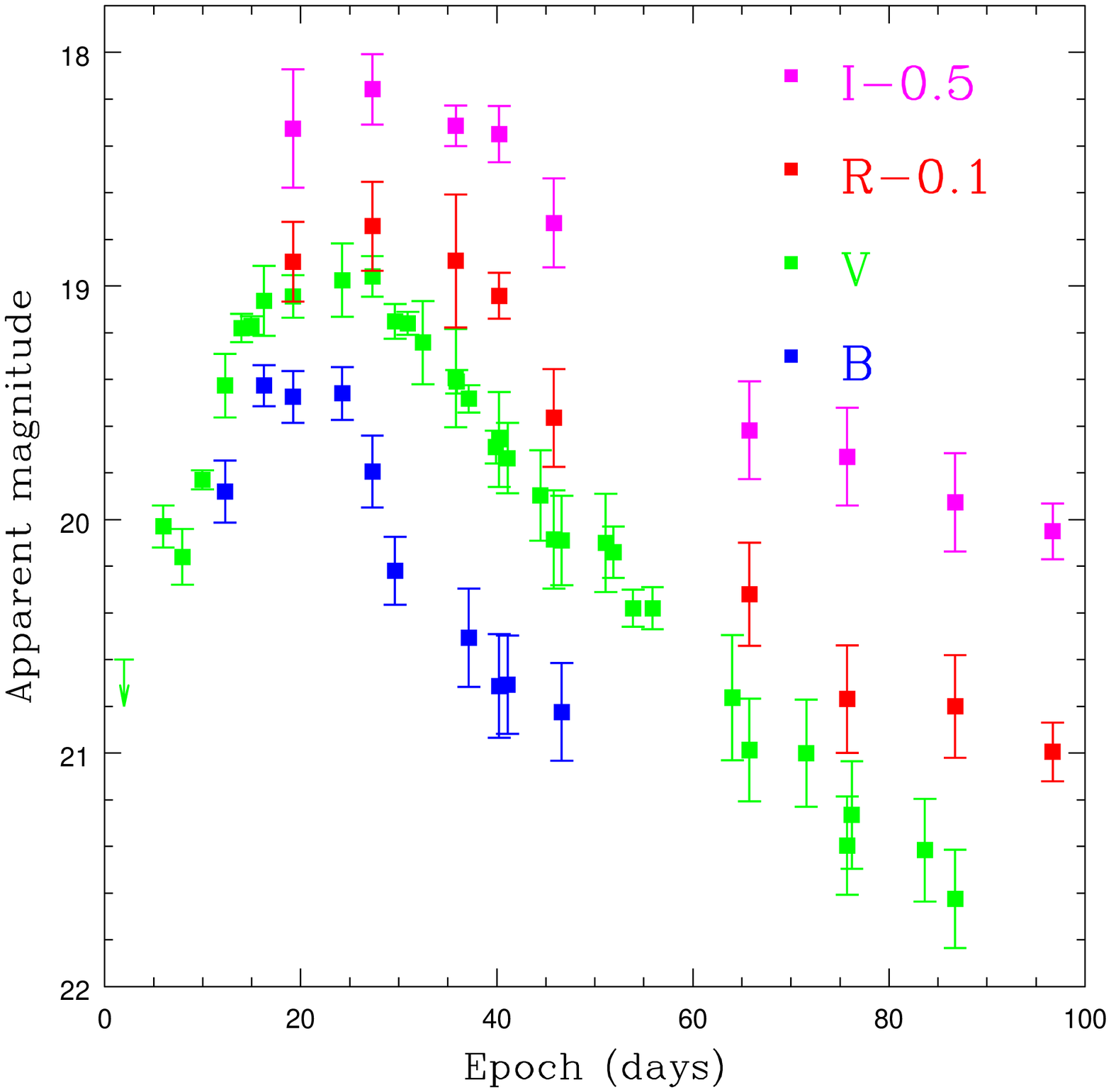}
  \includegraphics[scale=0.35]{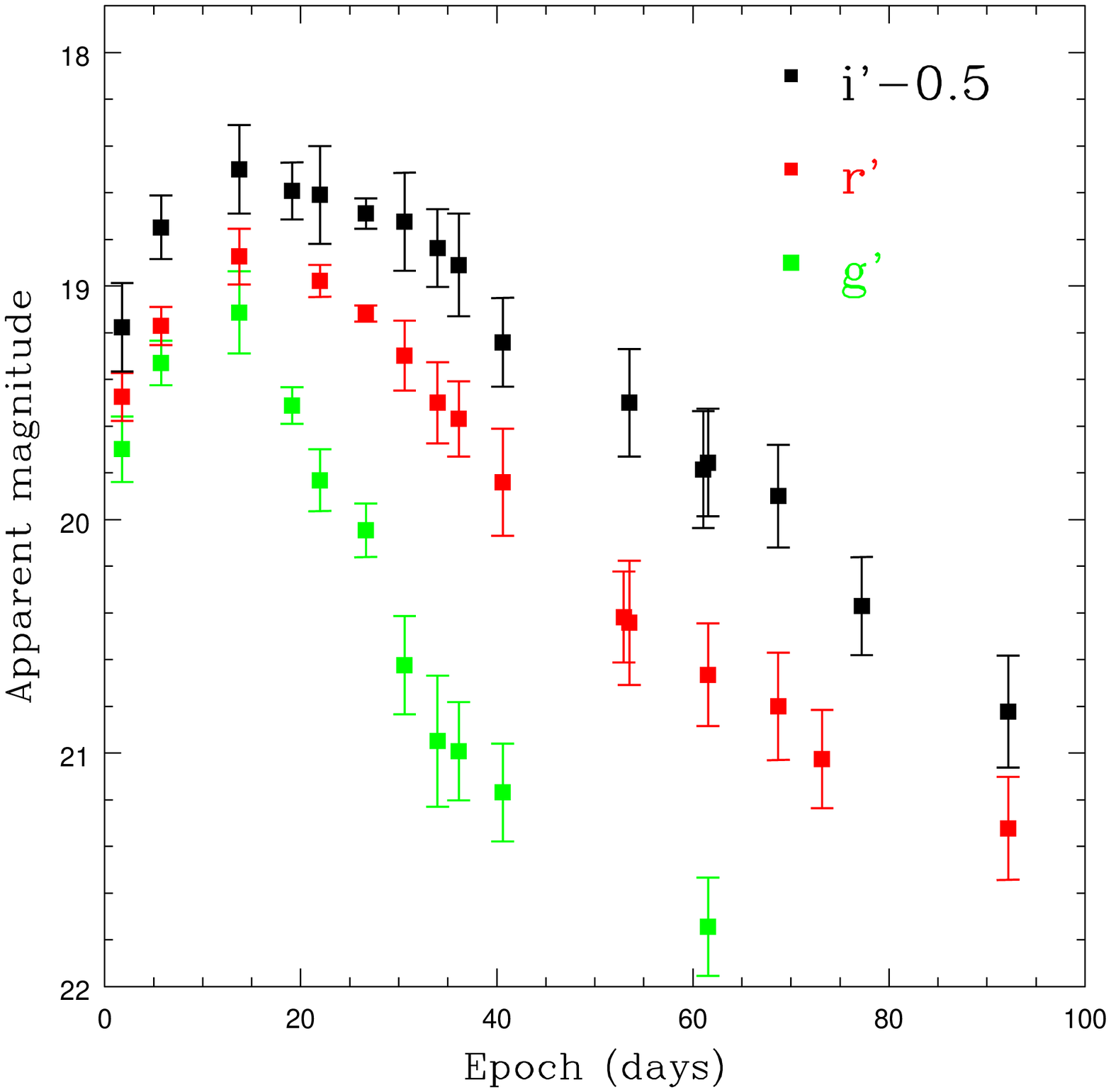}
  \caption{Left panel: photometric evolution of LSQ14efd in the Johnson-Cousins $BVRI$ filters, reported in Vega magnitude. Right panel: photometric evolution of LSQ14efd in the $g'r'i'$ filters, in AB magnitude. The pre-explosion limit in $V-band$ is represented with the arrow. Phases refer to the date of explosion. A shift has been applied for clarity.}
  \label{LC_opt}
\end{figure*}

The photometric evolution of LSQ14efd in the $BVRI$ and in the $g'r'i'$ filter systems is shown in Fig. \ref{LC_opt}.
The SN was discovered $14$ days before maximum in $B$ band.
The epoch of the B-maximum ($MJD=56895.72$) was obtained with a polynomial fit performed using the first 20 days of data. All subsequent
 epochs referred to in this work, unless specified, will refer to this date as the epoch zero. 

The B-band light curve reaches a peak magnitude of $m_{B}=19.47$ mag and has a decline rate of $5.61 \pm 0.78$ mag per 100 days, in the interval 5-30 days past B-maximum.
This interval was assumed to measure the decline rate also for all other bands.
The $V$ band light curve reaches the peak $\sim 5$ days after with a magnitude of $m_{V}=18.97$ mag and a decline rate of $4.48 \pm 0.45$ mag per 100 days.
In the $R$ band, the light curve peaks around $7$ days with a value of $m_{R}=18.84$ mag and a decline rate of $3.28 \pm 0.33$ mag per 100 days.
The $I$ band, the light curve peak appears around $8$ days with a magnitude of  $m_{I}=18.66$ mag and a decline rate of $2.28 \pm 0.23$ mag per 100 days.
In the $g'$ we see the peak after around $5$ days with a magnitude of $m_{g'}=19.11$ mag and a decline rate of $5.96 \pm 0.72$ mag per 100 days. 
The $r'$ band shows a peak after $\sim 5$ days with a magnitude of $m_{r'}=18.87$ mag and a decline rate of $2.78 \pm 0.28$ mag per 100 days.
The peak in the $i'$ band appears after around $5$ days showing a magnitude of $m_{i'}=18.96$ mag. The decline rate is $1.64 \pm 0.16$ mag per 100 days.
We also note that the blue bands show a narrower light curves compared to those of the red bands.

A comparison of the light curves of LSQ14efd with SN 2004aw \citep{Taubenberger2006} shows that the shift of the maximum in the different bands is compatible, within the errors, with the ones estimated for SN 2004aw, except for the maximum in the $V$ filter, which is reached after $\sim 3$ days for SN 2004aw and after $5$ days for LSQ14efd (see Table \ref{comp2004aw}).
A comparison of the decline rate, in the range 5-30 days, shows good agreement between the two SNe. However in the $B$ band  the decline rate measured for SN 2004aw it is $6.96 \pm 0.16$ mag per 100 days and $5.61 \pm 0.78$ mag per 100 days for LSQ14efd (see Fig. 2 in \citet{Taubenberger2006}).
This decline rate is slower than of SN 1994I which shows a faster evolution and a decline rate of $\sim 9$ mag per 100 days in the B band \citep{Elmhamdi2006}.

\begin{table*}
\caption{Comparison of LSQ14efd some photometric parameters with those of SN 2004aw.}\label{comp2004aw}
\begin{footnotesize}
\begin{tabular}{llcccc}
\hline
& & B & V & R & I  \\
\hline
Apparent magnitude at maximum & LSQ14efd & $19.47 \pm 0.11 $ & $18.97 \pm 0.15 $ & $18.84 \pm 0.19 $ & $ 18.66 \pm 0.15 $ \\
& SN 2004aw & $18.06 \pm 0.04$ & $17.30 \pm 0.03$ & $16.90 \pm 0.03$ & $16.53 \pm 0.03$ \\
Absolute magnitude at maximum & LSQ14efd$^{b}$ & $-18.04 \pm 0.18$ & $-18.50 \pm 0.21$ & $-18.60 \pm 0.24$ & $ -18.75 \pm 0.21$\\
& SN 2004aw$^{c}$ & $-17.63 \pm 0.48$ & $-18.02 \pm 0.39$ & $-18.14 \pm 0.34$ & $-18.18 \pm 0.28$ \\
Epoch of maximum$^{a}$ & LSQ14efd & $0.0$ & $+5.0 \pm 0.5$ & $+ 7.0 \pm 0.5$ & $+ 8.0 \pm 0.6$ \\
& SN 2004aw & $0.0$ & $+2,7 \pm 0.6$ & $+6.6 \pm 0.6$ & $+8.9 \pm 0.9$ \\
Decline rate$^{d}$ & LSQ14efd & $5.61 \pm 0.78$ & $4.48 \pm 0.45$ & $3.28 \pm 0.33$ & $2.28 \pm 0.23$ \\
& SN 2004aw & $6.96 \pm 0.16$ & $4.64 \pm 0.12$ & $3.20 \pm 0.12$ & $2.16 \pm 0.12$ \\
$\Delta m_{15}$ & LSQ14efd & $0.9 \pm 0.3$ & $0.3 \pm 0.1 $ & $0.1 \pm 0.1$ & $0.02 \pm 0.02$ \\
\hline
\end{tabular}
\\[1.5ex]
$^{a}$ Epoch relative to the B-maximum; 
$^{b}$ a distance modulus of $\mu = 37.35 \pm  0.15$ mag and a extinction $E(B-V) = 0.0376 \pm 0.0015$ mag were adopted for LSQ14efd; 
$^{c}$ a distance modulus of $\mu = 34.17 \pm  0.23$ mag and a extinction $E(B-V) = 0.37 \pm 0.10$ mag were adopted for SN 2004aw \citep{Taubenberger2006}; 
$^{d}$ decline rate in the time range $5-30$ days after B-maximum, in mag per 100 days.
\end{footnotesize}
\end{table*}

\begin{figure}
\centering
\includegraphics[scale=0.45]{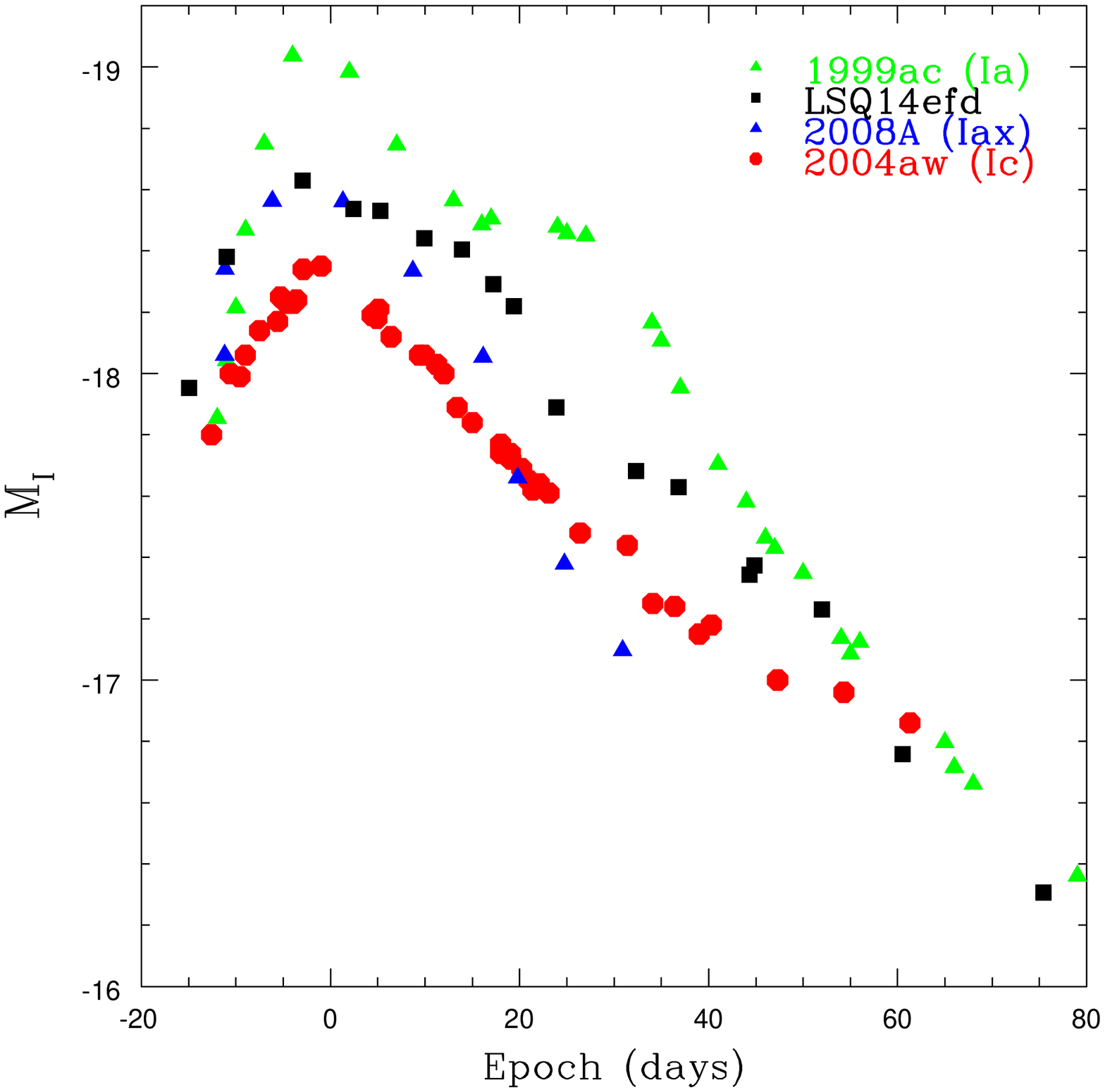}
\caption{Absolute I magnitude of LSQ14efd compared with those of SN-Ic 2004aw \citep{Taubenberger2006}, SN-Ia 1999ac \citep{Phillips2006} and SN-Iax 2008A \citep{McCully2014}. Epochs refer to the B-maximum.}
\label{I}
\end{figure}

As it will be shown later, the spectroscopic evolution of LSQ14efd shows some similarities with type Iax SNe, therefore it is useful to compare its photometric evolution with Iax and peculiar type Ia SNe to determine if there are
any similarities.  In Fig. \ref{I}, we present the absolute $I$ magnitude light curve of LSQ14efd compared with SNe 1999ac (normal Ia), SN  2008A (SN-Iax) and SN 2004aw (SN-Ic as discussed above). 
LSQ14efd does not show a secondary I-band peak as we see normal 
SN-Ia and it has a wider peak and a slower post-maximum decline
when compared the   type Iax SN 2008A. It bears 
most similarities with SN 2004aw, although the comparison is qualitative. 
The measure of the $\Delta m_{15}$ for  LSQ14efd is reported in Table \ref{comp2004aw}
We have applied the values of $\Delta m_{15}$ of the Phillip's relation \citep{Phillips1993} for type Ia SNe to LSQ14efd 
and  found an implied $M_{B}= -19.3 \pm 0.3$ mag.  This differs by $\sim 1.26$ mag from the measured B-band peak of SN  (see Tab. 3) which implys that LSQ14efd does not satisfy the SN-Ia relation and does not comfortably fit with this physical explanation. 
We stress that the type Iax SNe also do not satisfy the Philipps relation.

\begin{figure}
	\centering
	\includegraphics[scale=0.4]{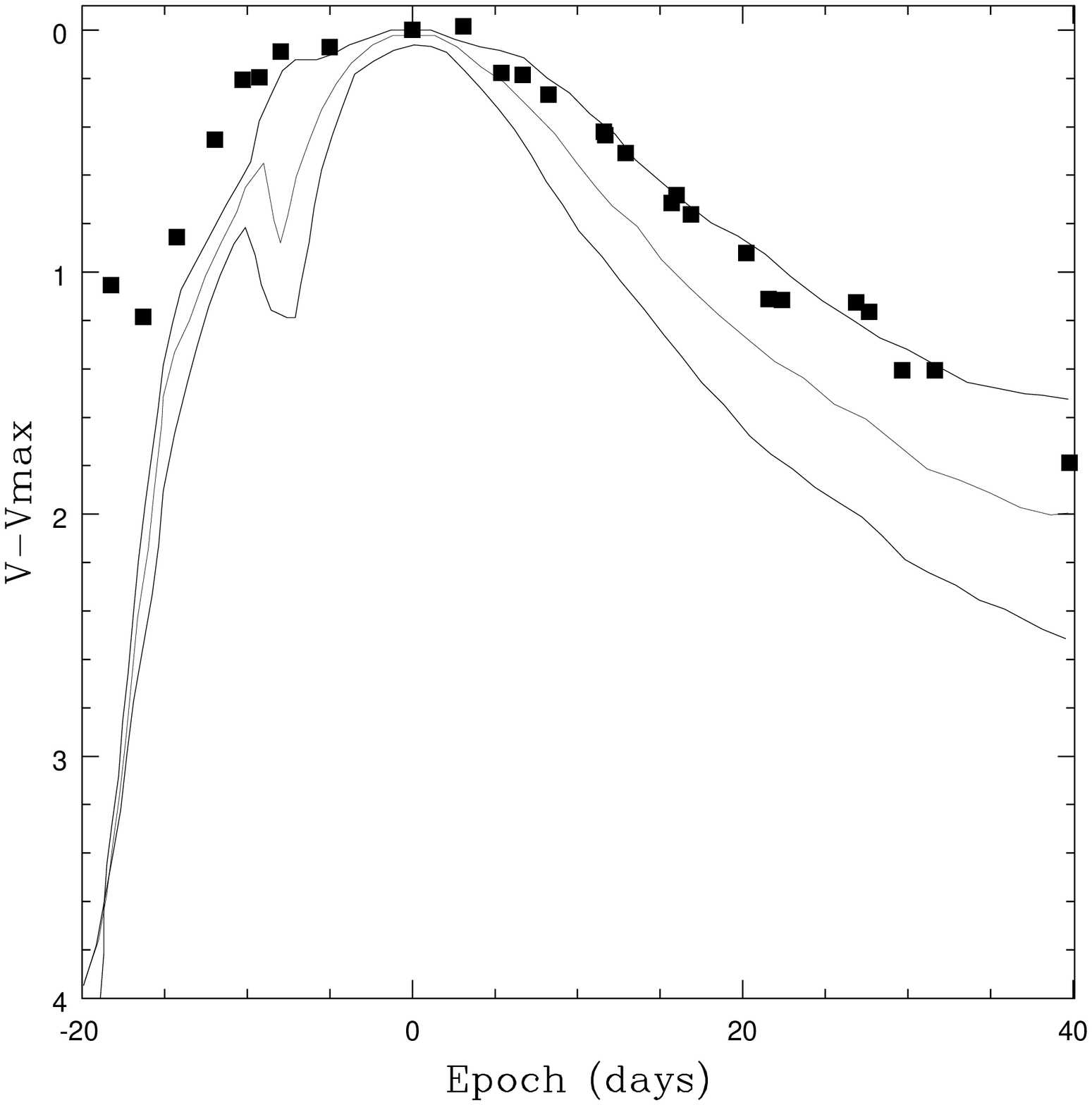} \\
	\includegraphics[scale=0.4]{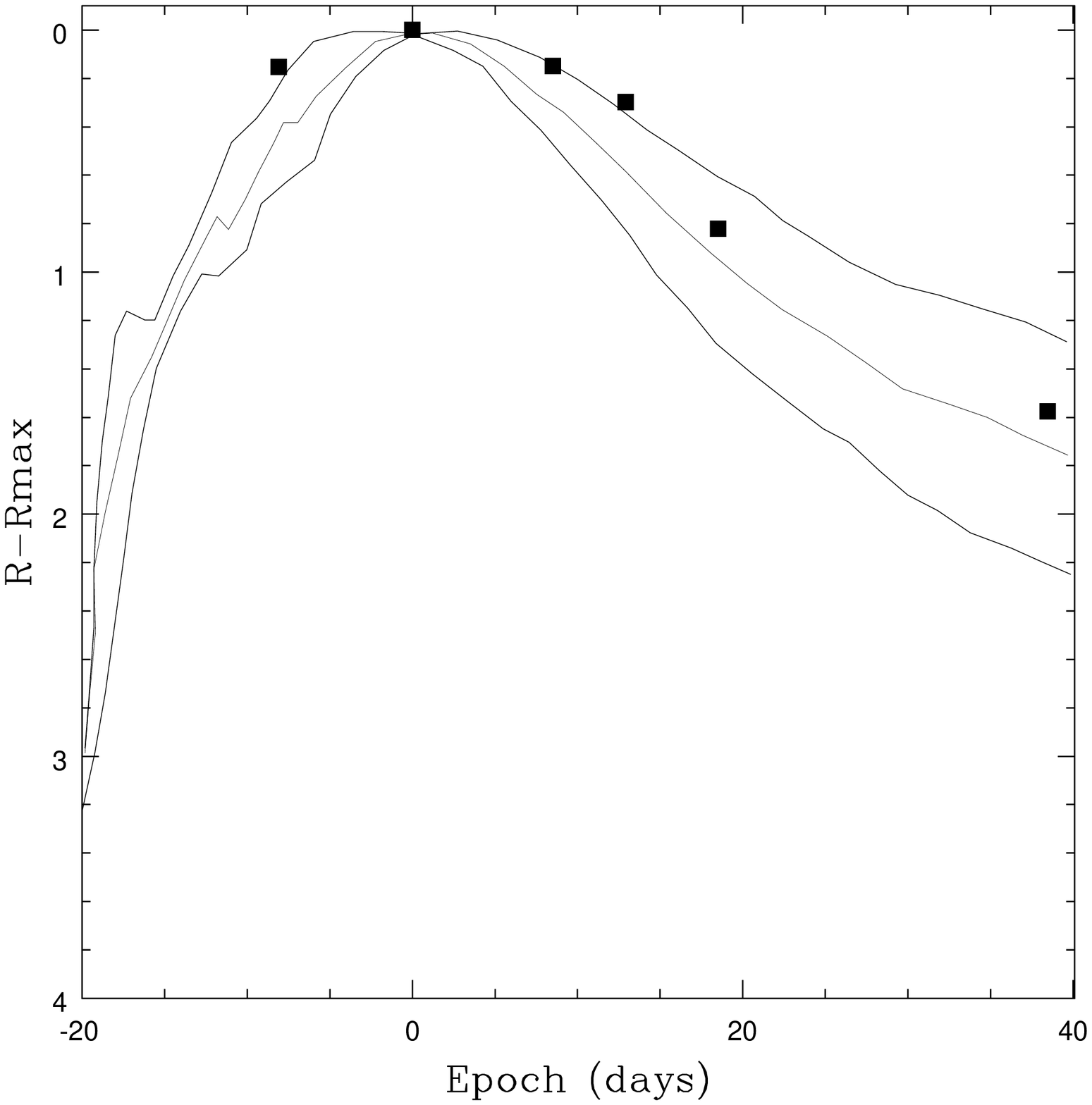}
	\caption{Comparison of V and R band light curve of LSQ14efd and the templates found in \citealt{Drout2011} for SNe-Ibc. The central line shows the best fit while the two outer lines show the 1$\sigma$ contours. Epochs refers to the maximum of the corrispondent band.
	}
	\label{R_template}
\end{figure}

We then compared the light curves of LSQ14efd with those of a sample of SNe-Ib/c. In particular, the evolution of the light curve in the R and V bands of LSQ14efd was compared with the templates by \citet{Drout2011} (Fig. \ref{R_template}). Those template are the result of the interpolation over the normalized V and R light curve of 10 well-sampled literature SNe-Ib/c. The weighted mean flux density was then extracted over the time interval -20 to 40 days. We notice that the evolution of the V band of LSQ14efd follows the decline post-maximum of the template within $1\sigma$ while the pre-maximum evolution differs significantly from the template showing that the light curve of LSQ14efd is broader than the ones in the sample. Instead, the R band evolution of LSQ14efd follows nicely the template but we note that in the R band we are missing data at early phases when the most significant deviation is observed in the V-band.

\begin{figure*}
  \centering
  $
  \begin{array}{c@{\hspace{.1in}}c@{\hspace{.1in}}c}
  \includegraphics[scale=0.63]{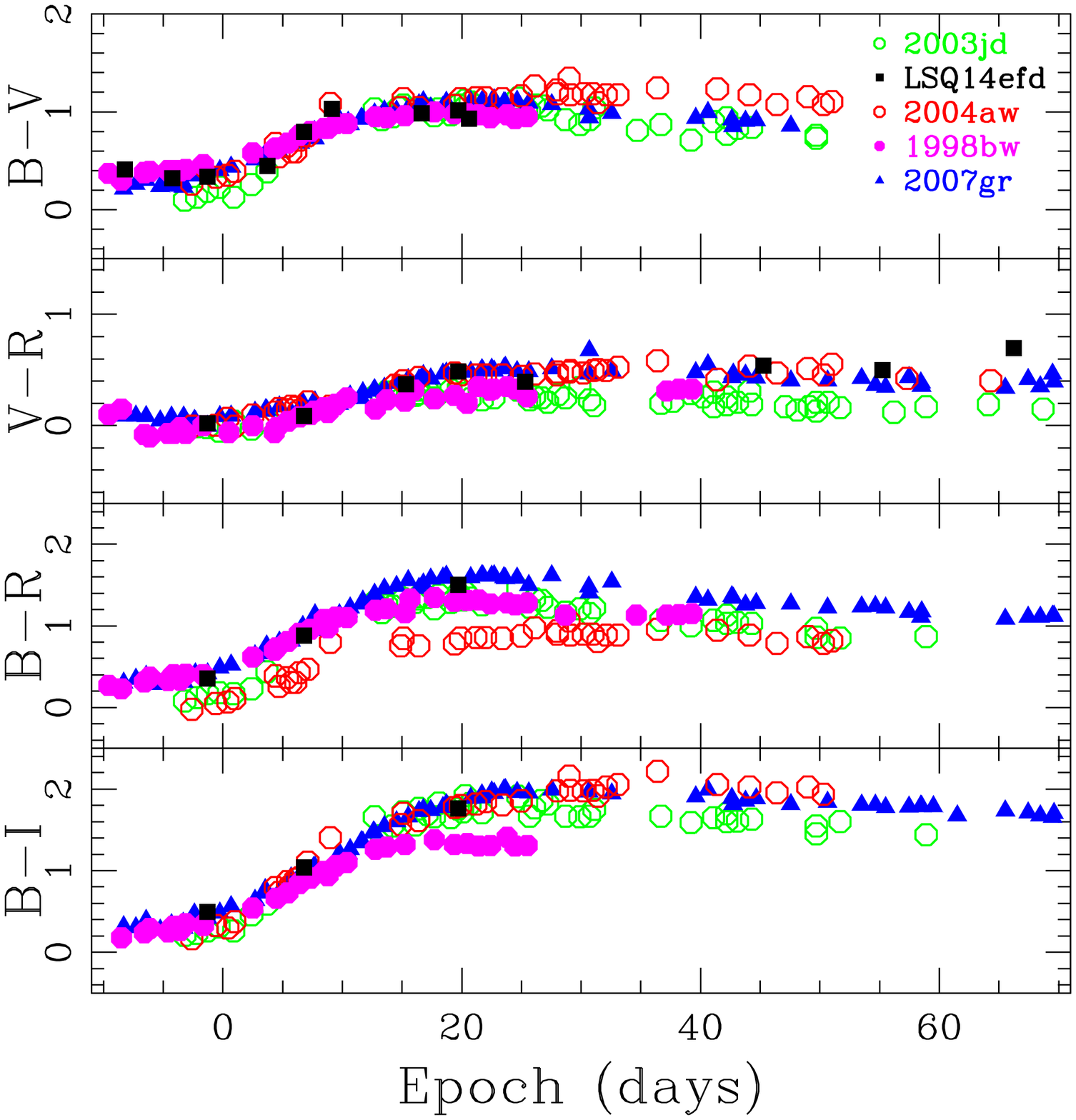}\\
  \includegraphics[scale=0.63]{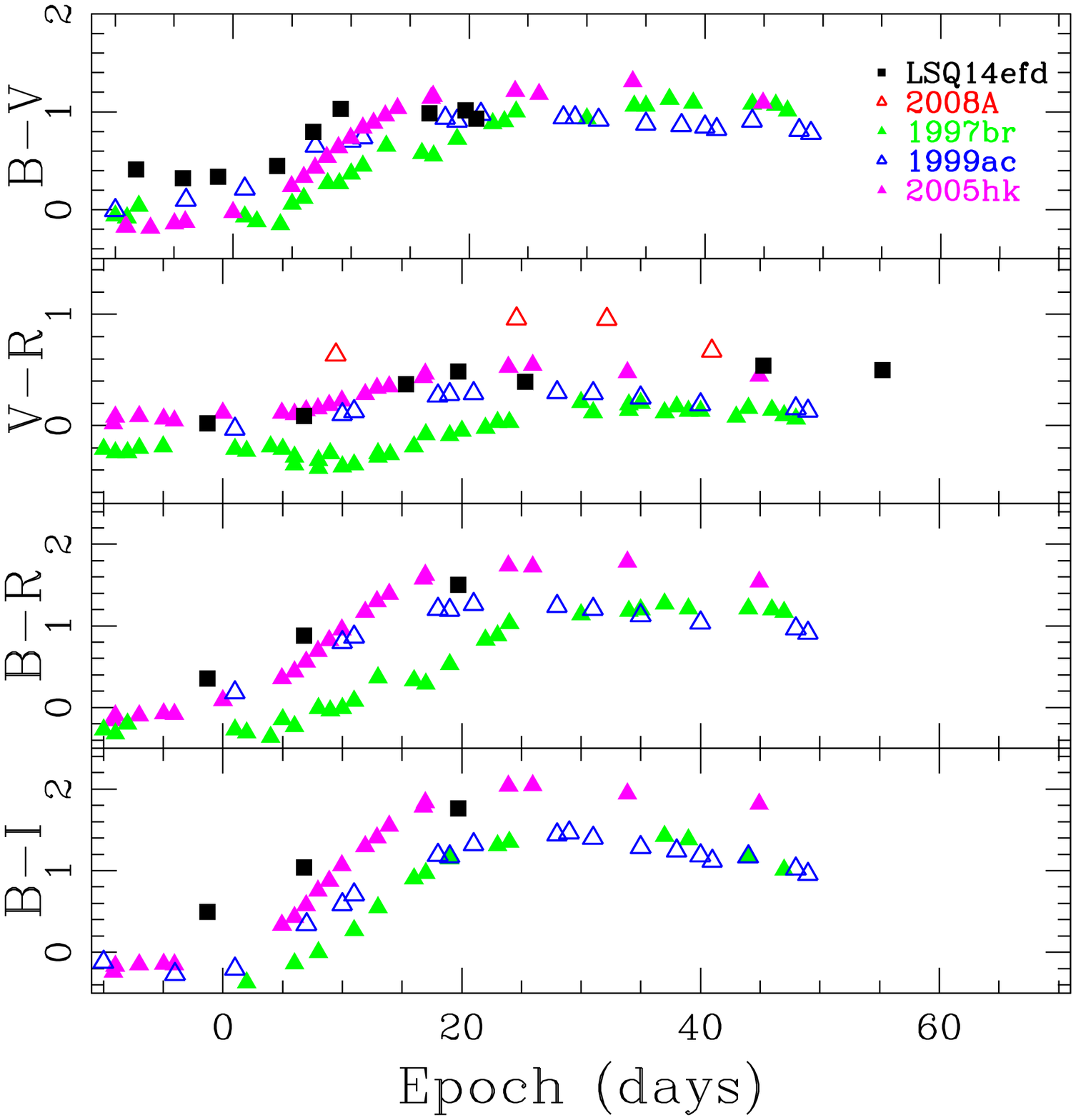}
  \end{array}
  $
  \caption{Top panel: colour evolution of LSQ14efd compared to those of SNe-Ic, 2004aw \citep{Taubenberger2006}, 2003jd \citep{Valenti2008a}, 2007gr \citep{Valenti2008b} and 1998bw \citep{Patat2001}. Bottom panel: colour evolution of LSQ14efd compared to those type Ia SNe, 1999ac \citep{Phillips2006}, 1997br \citep{Li1999} and type Iax SN 2008A \citep{Foley2013} and 2005hk \citep{Phillips2007}. Epochs refer to the B-maximum.}
  \label{color_all}
\end{figure*}

The dereddened $B-V$, $V-R$, $B-R$ and $B-I$ colour evolution are shown in Fig \ref{color_all}.
We measure a $B-V$ colour of $0.4$ mag at $\sim 10$ days before B-maximum.
It  then increases to about 1 mag at $\sim 5$ days and stays more or less constant around this value. The $V-R$ colour increases from $\sim 0$ mag to around $0.5$ mag within $15$ days after B-maximum and settles at around this value for the subsequent days.
The $B-R$ colour steadily increases from $\sim 0.3$ mag to about $1.4$ mag within $\sim 20$ days after B-maximum. The $B-I$ colour increases from $\sim 0.5$ mag to around $1.7$ mag at $\sim 20$ days after B maximum.

The dereddened colour evolution have been compared with those of some type Ic SNe (2004aw, \citealt{Taubenberger2006}, 2003jd \citealt{Valenti2008a}, 2007gr \citealt{Valenti2008b,Hunter2009} and 1998bw \citealt{Patat2001}), to those of some type Ia SNe (1999ac \citealt{Phillips2006}, 1997br \citealt{Li1999}) and to those of some type Iax SNe (2008A \citealt{Foley2013} and 2005hk \citealt{Phillips2007}).

The colour evolutions of type Ic SNe are well defined in the $B-V$, $V-R$ and $B-I$ colours (see Fig. \ref{color_all}, upper panel). In the $V-R$ and $B-I$ evolution we can see that SN 1998bw shows a slightly different trend after $\sim 15$ days, with a flattening in the evolution that is shown only $\sim 10$ days after in the other SNe of this sample.
The trend of the $B-R$ evolution is also quite similar for all the SNe of this sample. We notice that the colour evolution of LSQ14efd is very similar to those of type Ic SNe. They show fairly similar trends in the rising part of curves and they all subsequently  flatten to comparable values.
The similarity between dereddened colours of LSQ14efd and those of the type Ic comparison sample supports the hypothesis of no extinction within LSQ14efd host galaxy (Sect. \ref{discovery}).

The $B-V$, $B-R$ and $B-I$ colour evolutions of type Ia and Iax SNe also show a broadly similar trend. The $V-R$ colour evolution 
is  the most diverse among the objects of the sample (see Fig. \ref{color_all}, lower panel).
In $V-R$ 
the colour evolution of LSQ14efd is similar to that of SN 2005hk while SN 2008A appears to diverge from the broad trends of the set.
The $B-V$, $B-R$ and $B-I$ colour evolutions of LSQ14efd seems to be bluer at early phases and around maximum with respect to those of SNe 2005hk and 1999ac. In the $B-R$ and $B-I$ colour evolutions we point out that the curves are
steeper for Ia and Iax SNe respect to those of LSQ14efd.

\subsection{Quasi-bolometric light curve}\label{bolometric}

\begin{figure*}
  \centering
  \includegraphics[scale=0.5]{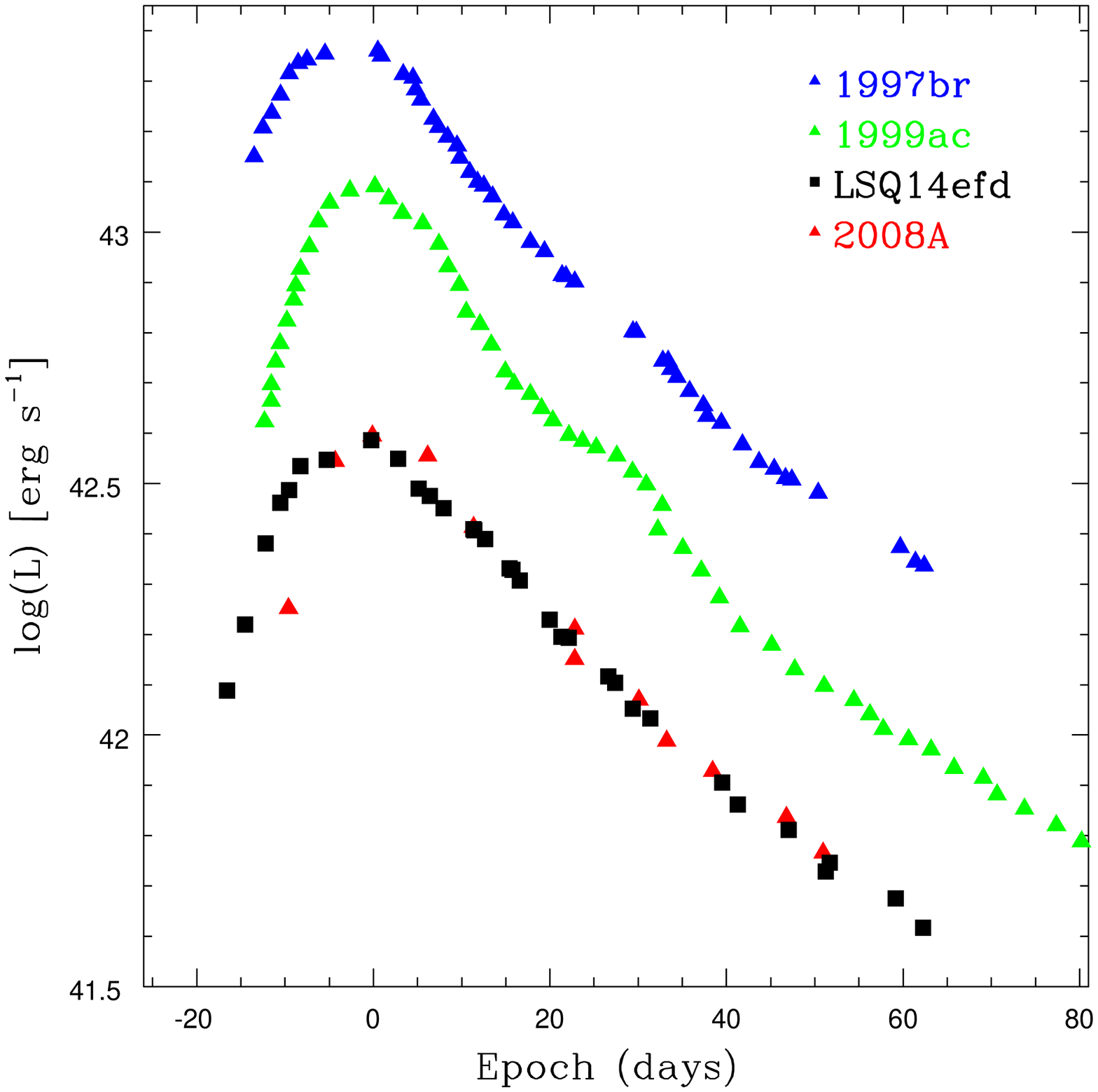} \\
  \includegraphics[scale=0.5]{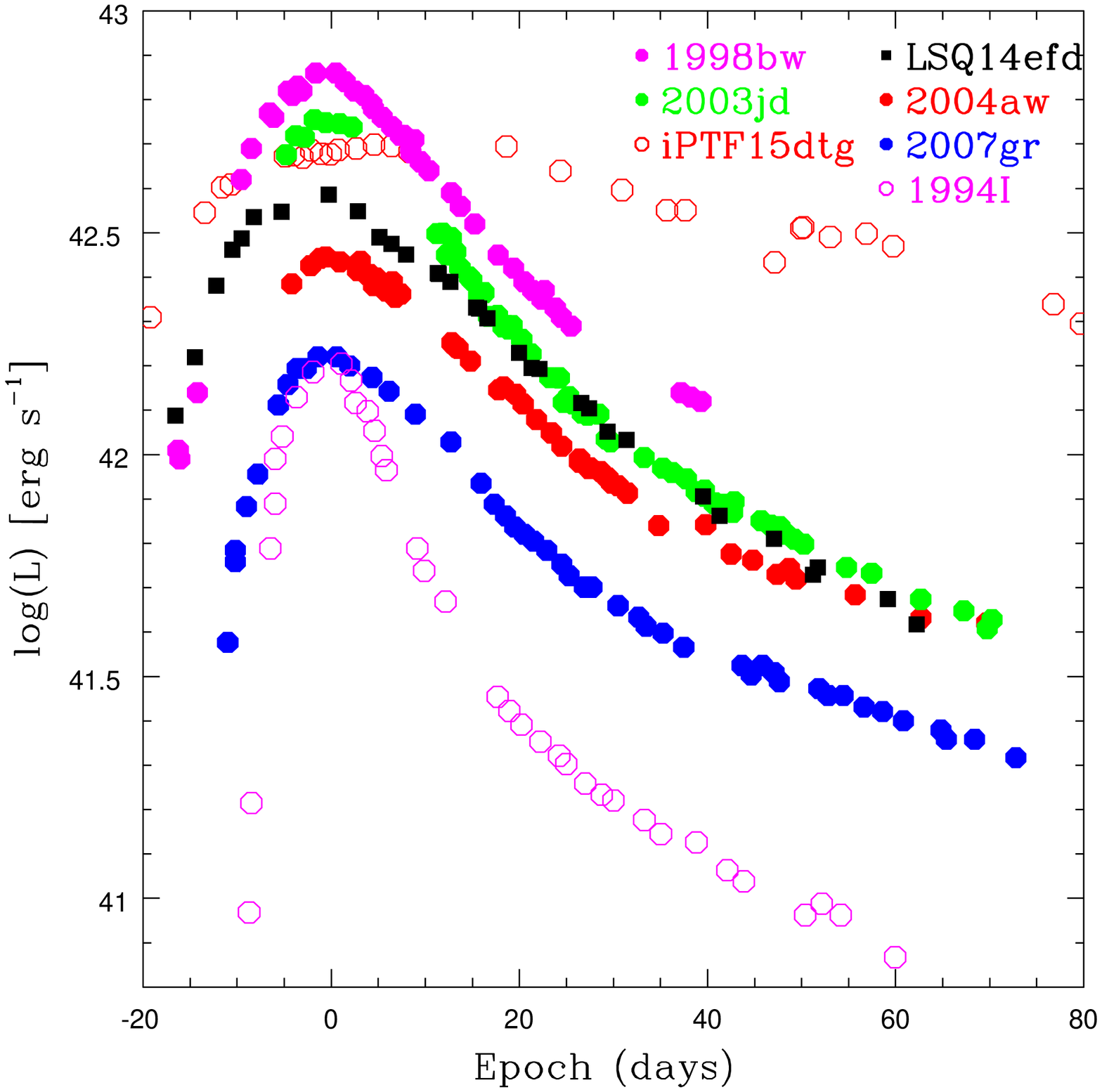}
  \caption{\textit{Upper panel}: Quasi-bolometric light curve of LSQ14efd compared with type Ia SNe 1999ac \citep{Phillips2006} and 1997br \citep{Li1999} and with type Iax SN 2008A \citep{McCully2014}. \textit{Lower panel}: Quasi-bolometric light curve of LSQ14efd compared with type Ic SNe 2007gr ($UBVRI$, \citealt{Valenti2008b}), 2004aw ($UBVRI$, \citealt{Taubenberger2006}), 1994I ($UBVRI$, \citealt{Richmond1996}), 2003jd ($BVRI$, \citealt{Valenti2008a}), iPTF15dtg ($gri$, \citealt{Taddia2016}) and 1998bw ($UBVRI$, \citealt{Patat2001}. Phases are respect to the B-maximum of each SN.}
  \label{bolom_all}
\end{figure*}

A quasi-bolometric light curve has been calculated by integrating the 
observed optical flux over wavelength.
The estimated $Bg'Vr'Ri'I$ apparent magnitudes were converted into monochromatic fluxes at the effective wavelength for each filter. After correcting for Galactic extinction (Sect. \ref{discovery}), the resulting Spectral Energy Distribution (SED) has been integrated over the full observed wavelength range, assuming, at limits, a zero flux. The flux was estimated at the phases in which V band observations were available. For the other bands for which photometry was not available, the magnitudes were calculated by interpolation of the values from the photometry on nearby nights.
Finally, using the redshift-based distance of the galaxy (Sect. \ref{discovery}), the integrated fluxes were converted into luminosity.
We note that the first point was excluded when building the quasi-bolometric light curve.
The quasi-bolometric light curve of LSQ14efd is shown in Fig. \ref{bolom_all} together with some other type Ia and Ic SNe.

The luminosity at the peak is $L = 3.9 \times 10^{42} \rm \: erg \: s^{-1}$. This has to be considered as a lower limit since we constructed the quasi-bolometric light curve only across the wavelength limits sampled by the
observed filter set, namely the $Bg'Vr'Ri'I$ bands. \citet{Lyman2014} developed a method to estimate the bolometric correction from data which 
are limited in wavelength. The value of the peak bolometric luminosity estimated with this method is $L \sim 5 \times 10^{42} \rm \: erg \: s^{-1} $. \\
In Fig \ref{bolom_all}, we can notice that the post-max slope of the light curve of LSQ14efd compares well with those of type Ic SNe 1998bw, 2003jd, iPTF15dtg and 2004aw, while its peak luminosity is intermediate between 2003jd and 2004aw. It is significantly less luminous than the SN-Ic-BL 1998bw but with a similar width. LSQ14efd shows a different behaviour from type Ic SNe 1994I and 2007gr, being more luminous and showing a broader bolometric curve. Instead, iPTF15dtg has a comparable luminosity with respect to LSQ14efd but it is much wider and show a slower decay. We note that type Ia SN 1999ac shows an evident double peak in the curve, which is not present in the light curves of the other SNe plotted. 
Also the peak luminosity  of type Ia SN 1999ac is much brigther than LSQ14efd. SN-Ia 1997br has a comparable light curve shape with respect to LSQ14efd but it is more luminous at every epoch. The peak luminosity of LSQ14efd is not far from  that 
of the type Iax SN 2008A, but it shows a wider peak and a slower decay.
We can conclude from this comparison that the photometric evolution of LSQ14efd is consistent with those of the known population of
type Ic SNe but also similar to peculiar type Iax SNe.

\section{Spectroscopic evolution}

\subsection{Data sample and reduction}\label{Specsec}

We performed a spectroscopic monitoring campaign of  LSQ14efd at the ESO NTT at La Silla, Chile. Eight epochs of optical spectra were acquired with  EFOSC2 from $8$ days before B-maximum until $37$ days after B-maximum.
Details of the spectroscopic observations and the characteristics of the employed instrumentation are listed in Table \ref{logspec}.

\begin{table}
\setlength{\tabcolsep}{0.5pt}.
\caption{Summary of instrumental sets-up used for the spectroscopic follow-up campaign.}\label{logspec}
\begin{footnotesize}
\begin{tabular}{lcllcc}
\hline
Telescope & Instrument & Grism & Range & Resolution & \# of epochs  \\
 &  & & [ \AA\ ] & [ \AA\ ] & \\
\hline
NTT (3.58m)    & EFOSC2 & Gr13  & 3985-9315 & 18 & 7 \\
NTT (3.58m)    & EFOSC2 & Gr11  & 3380-7520 & 12 & 1 \\
\hline
\end{tabular}
\\[1.5ex]
NTT = New Technology Telescope with the optical camera ESO Faint Object Spectrograph and Camera EFOSC2

\end{footnotesize}
\end{table}

\begin{figure}
	\centering
	\includegraphics[scale=0.4]{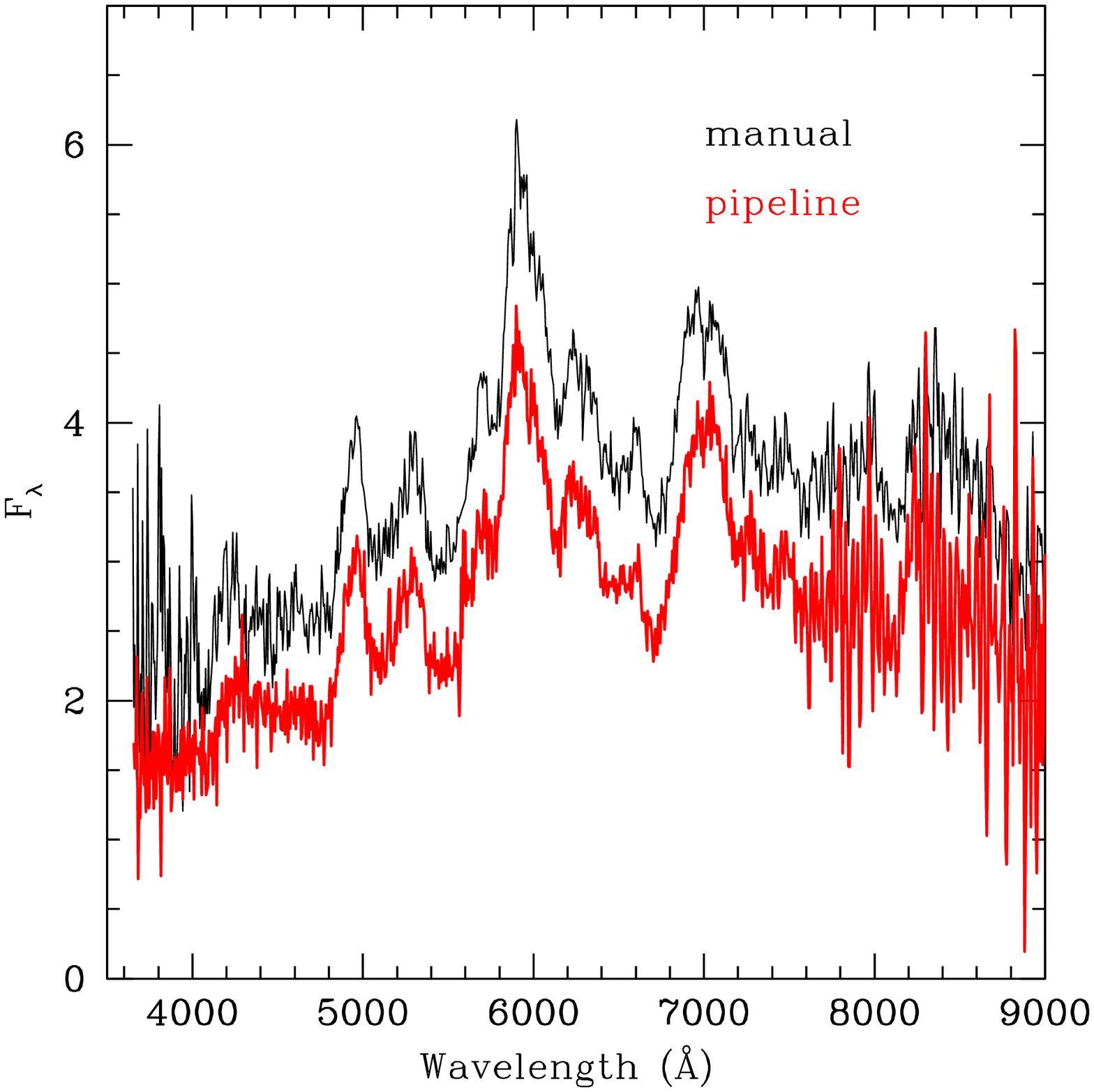}
	\caption{Comparison of the manually reduced spectrum of LSQ14efd at 28 days with the same obtained by the pipeline reduction.}
	\label{red_spec}
\end{figure}

The pre-reduction of the spectra (trimming, overscan, bias and flat-field correction) was performed using the PESSTO pipeline \citep{Smartt2015}, which is based on the standard IRAF tasks. Comparison spectra of arc lamps, obtained in the same instrumental configuration of the SN observations, were used for the wavelength calibration. Observations of spectrophotometric standard stars were used for the flux calibration. Atmospheric extinction corrections were applied using tabulated extinction coefficients of each telescope site.
We note that the data presented and analysed in this paper were custom re-reduced, and differ somewhat from those in formal public release of the PESSTO Spectral data products\footnote{The spectra will be available on WISeREP public database \citep{Yaron2012}, at http://wiserep.weizmann.ac.il.  All other data can be accessed also from the ESO Phase 3 archive, with all details on www.pessto.org}. We obtained some better quality results with more tailored and manual reductions, 
particularly with manual fringing corrections for the fainter spectra. In this case, the reduction of the spectra followed the standard procedure, with particular attention to the flat-field, considering just a few columns (100) next to the target, such to improve the removing of the fringing in the red bands. Also the background has been optimized, considering also an adjacent area, to minimize the contamination of the host galaxy.
An example of the difference in the data quality from the manual reduction and the pipeline is shown in Fig. \ref{red_spec}.

A comparison of synthetic $BV$ and $r$ photometry obtained from the spectra, throught the IRAF task \texttt{CALCPHOT}, with the observed photometry at similar epochs was performed to check the quality of the flux calibration. These spectro-photometric magnitudes were compared with those from photometric observations and, when required, a scaling factor was applied. Finally, calibrated spectra were dereddened for the total extinction and corrected for the estimated redshift.

\subsection{Data analysis}\label{specanalysis}

\begin{figure*}
  \includegraphics[scale=.7, angle=0]{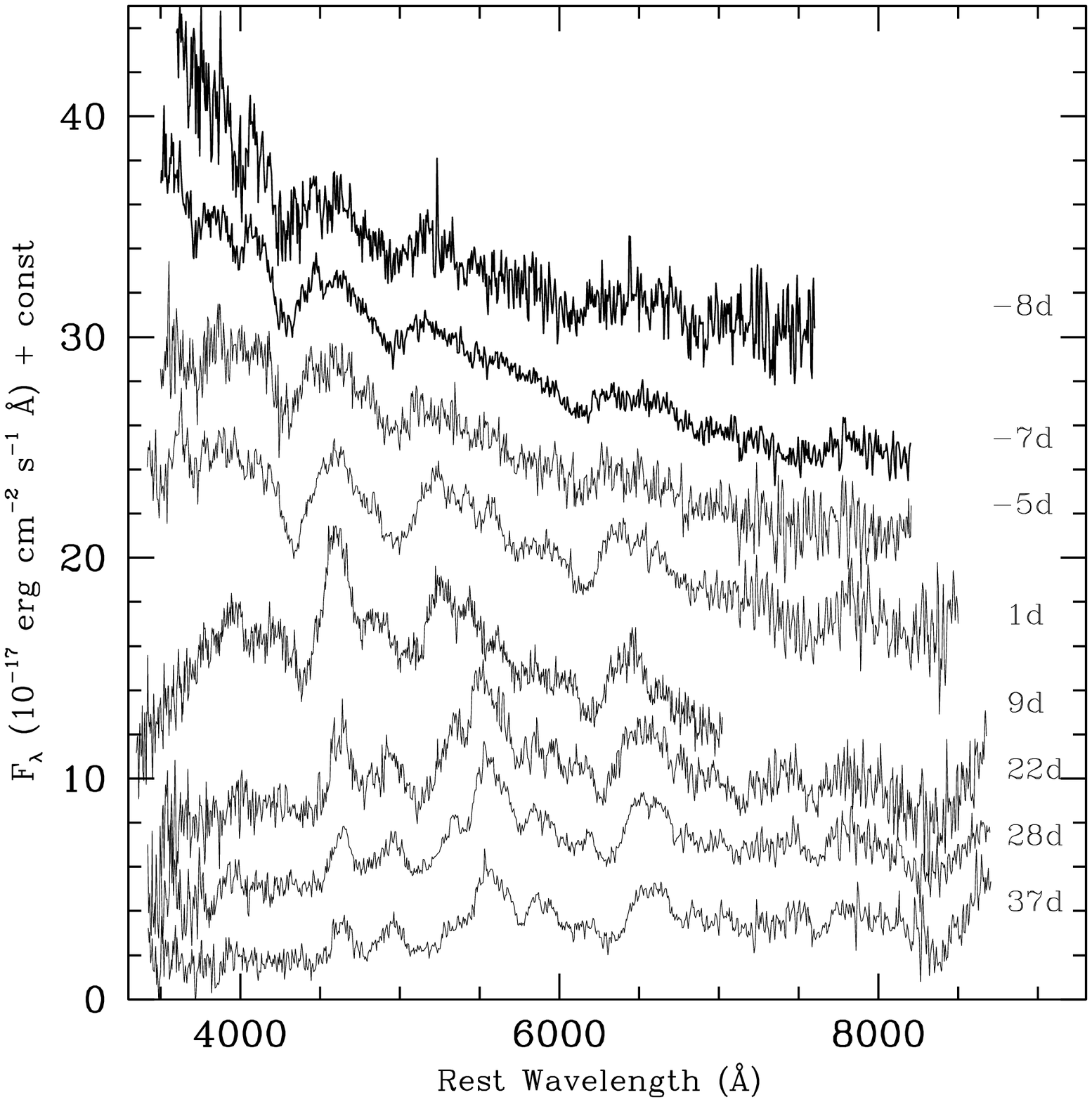}
  \caption{Optical spectroscopic evolution of LSQ14efd, starting from $-8$ days from the B-maximum to $37$ days after.}\label{spec_evol}
\end{figure*}

\begin{figure*}
  \centering
  \includegraphics[scale=.7, angle=0]{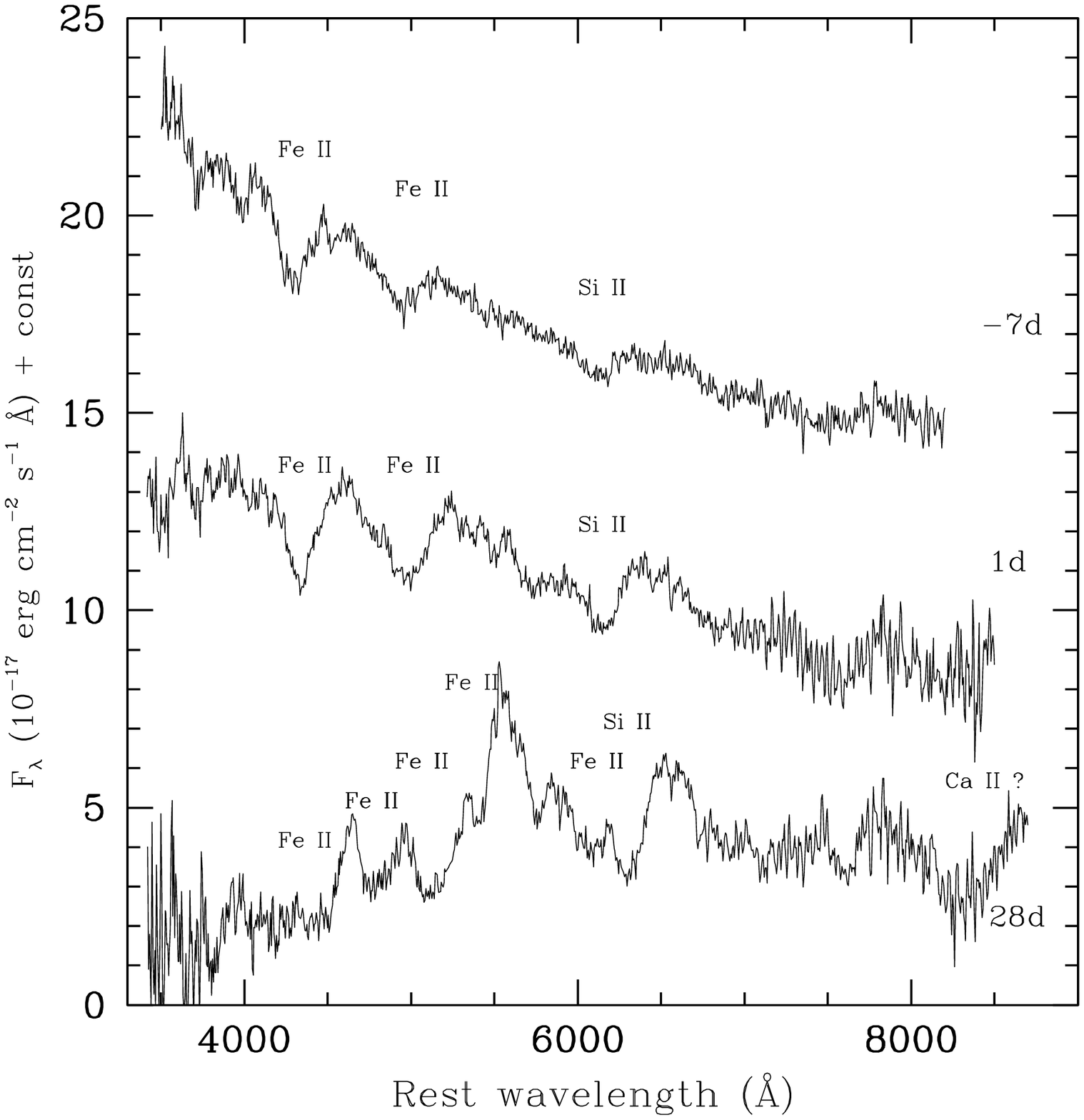}
  \caption{Identifications of line features observed in optical spectra of LSQ14efd, at three characteristic epochs.}\label{spec_id}
\end{figure*}

The time evolution of the optical spectra of  LSQ14efd, obtained from $-8$ to $37$ days with respect to B-maximum and covering the whole photospheric phase, is shown in Fig. \ref{spec_evol}. Corresponding line identifications are presented in Fig. \ref{spec_id}.

Spectra obtained one week before maximum show a blue continuum with some features due to Fe II lines ($\lambda\lambda\lambda$4440, 4555, 5169).
A feature very likely due to the Si II ($\lambda$6355) appears around phase -5. \citet{Branch2002} pointed out the  possibility of a misleading identification of the Si II in type Ib and Ic SNe as the H$\alpha$ line (see also \citealt{Parrent2015}) at high velocities (or ``detatched hydrogen"). 
As shown below in this section, the estimated velocities of Si II and Fe II are in agreement, within the errors, which suggests to us that
 identification as Si II  is the more plausible explanation. 
There is no evidence of narrow interstellar medium NaID absorption in the spectra, leading us to conclude the absorption due to the host galaxy is negligible, or at least not measurable with the data available. 
 During the evolution of the photospheric phase the spectra obtained close to maximum light show that the intensity of the blue continuum decreases significantly. It's also possible to see prominent features due to Fe II ($\lambda\lambda\lambda$4555, 5169, 5535) and the Si II line ($\lambda$6355). A possible feature due to Ca II ($\lambda 8542$) starts to appear.
Late spectra show a continuum dominated by the iron-group elements, we find Fe II ($\lambda\lambda$4555, 4924) and Fe II($\lambda\lambda$5169, 5535) while Si II lines ($\lambda$6355) are still barely visible.
There is no evidence of forbidden lines arising in the last spectrum, leading
 us to conclude the  LSQ14efd is still in the photospheric phase.

\begin{figure*}
  \includegraphics[scale=.7, angle=0]{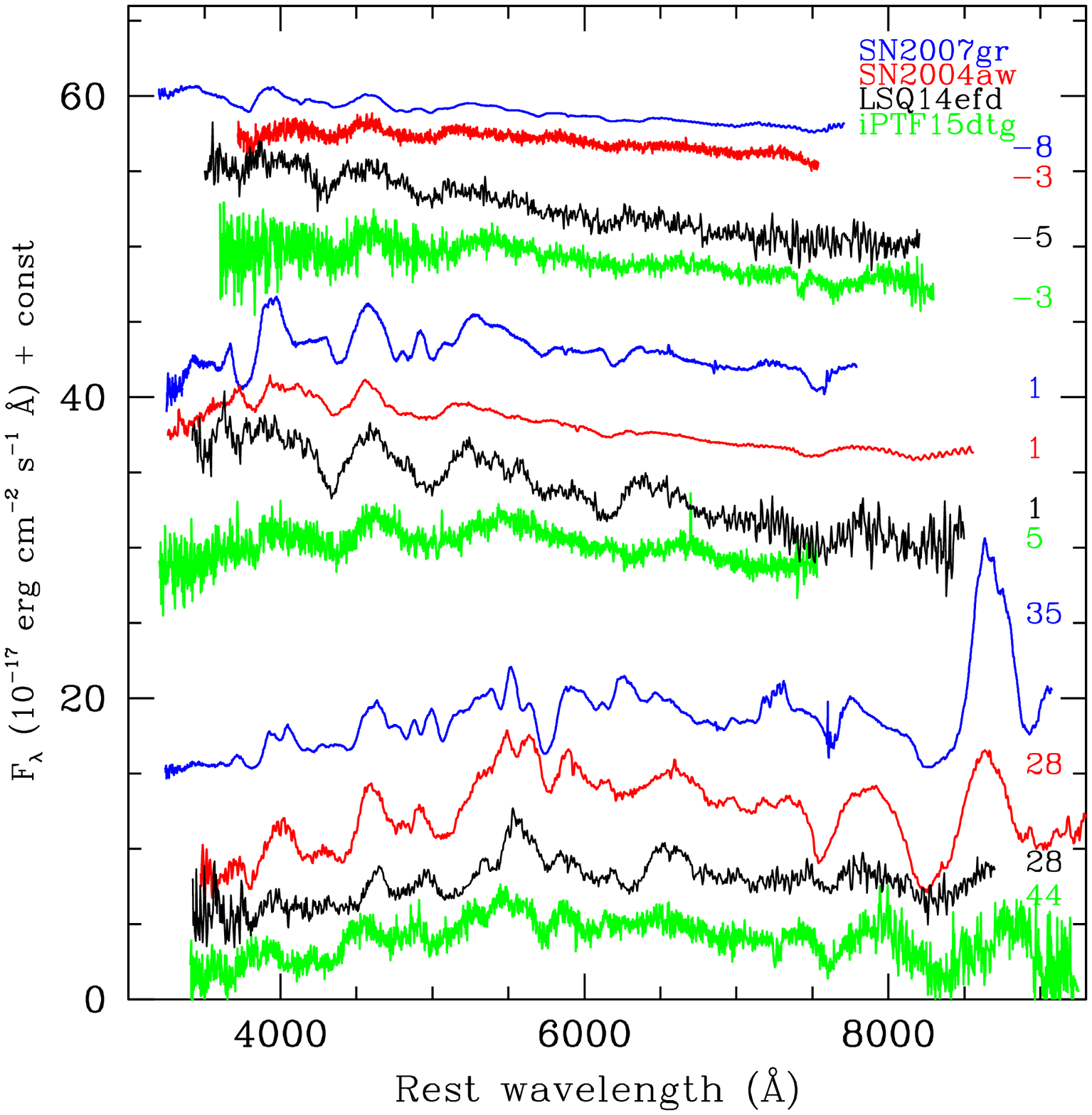}
  \caption{Comparison of spectra of LSQ14efd with those of type Ic SNe 2007gr \citep{Valenti2008b}, iPTF15dtg \citep{Taddia2016} and 2004aw \citep{Taubenberger2006}, at early pre-maximum epoch ($\sim -5$ days), around maximum and at late spectra ($\sim 28$ days)}\label{comp_all}
\end{figure*}

A comparison with other type Ic SNe is shown is Fig. \ref{comp_all}. 
The earliest LSQ14efd spectrum resembles the  featureless continuum of SNe 2007gr, iPTF15dtg and 2004aw (admittedly having low S/N). The spectrum around B-maximum does not show any major differences with those of SNe 2004aw and iPTF15dtg. SN 2007gr shows a generally good agreement, but with the presence of extra features around 4000 \AA\ and 5000 \AA. \\
The feature at around $8200$ \AA $\:$ is possibly due to Ca II, as the comparison with SN 2004aw seems to suggest. At the later stages, about 1 month past maximum, the spectrum of LSQ14efd exhibits similarities with iPTF15dtg, while there is a good agreement with SNe 2007gr and 2004aw, though they exibit  stronger Ca II and O I lines.

\begin{figure*}
\begin{center}
\includegraphics[scale=.7]{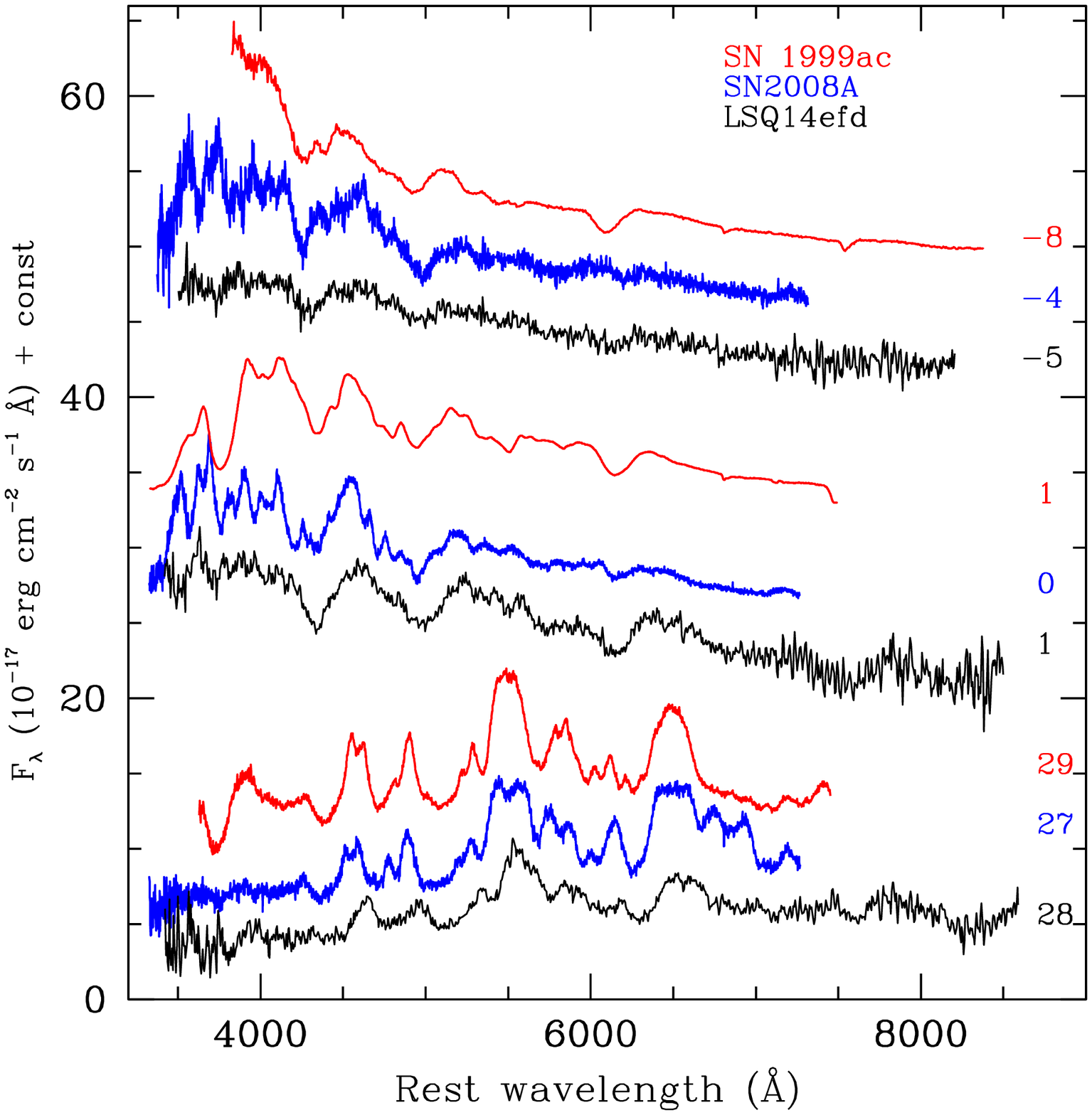}
\end{center}
\caption{Comparison of spectra of LSQ14efd with type Ia SN 1999ac \citep{Garavini2005}, type Iax SN 2008A \citep{Blondin2012}, at different epochs.}
\label{comp1999ac_08A}
\end{figure*}

Fig \ref{comp1999ac_08A} shows the early spectrum of  LSQ14efd 
(a few days before B-maximum) compared with  pre-maximum spectra of type Ia SN 1999ac and type Iax SN 2008A. In the blue part of the spectrum, LSQ14efd has lower signal to noise than that of SNe 1999ac and 2008A, while the rest resembles an almost featureless continuum. The LSQ14efd spectrum around maximum differs, from that of SNe 1999ac and 2008A, mainly in the blue part. The late spectrum of LSQ14efd is again similar to that of SNe 1999ac and 2008A, though the instensity of the Si II feature is weaker for LSQ14efd.
A comparison with SNe-Ic and SNe-Ic-BL templates \citep{Liu2016,Modjaz2014} remarks a quite good similarity with SNe-Ic for the early spectra, considering the width of the lines present in both, LSQ14efd and the template. We then note a significant deviation for the late ones since the strenght and the width of emission line of LSQ14efd differ from the those of the template. A similar comparison of LSQ14efd spectra with the template for SNe-Ic-BL seems to show also a good agreement at all epochs. This strengthens the peculiarity of LSQ14efd.

\begin{figure*}
	\begin{center}
		\includegraphics[scale=.6]{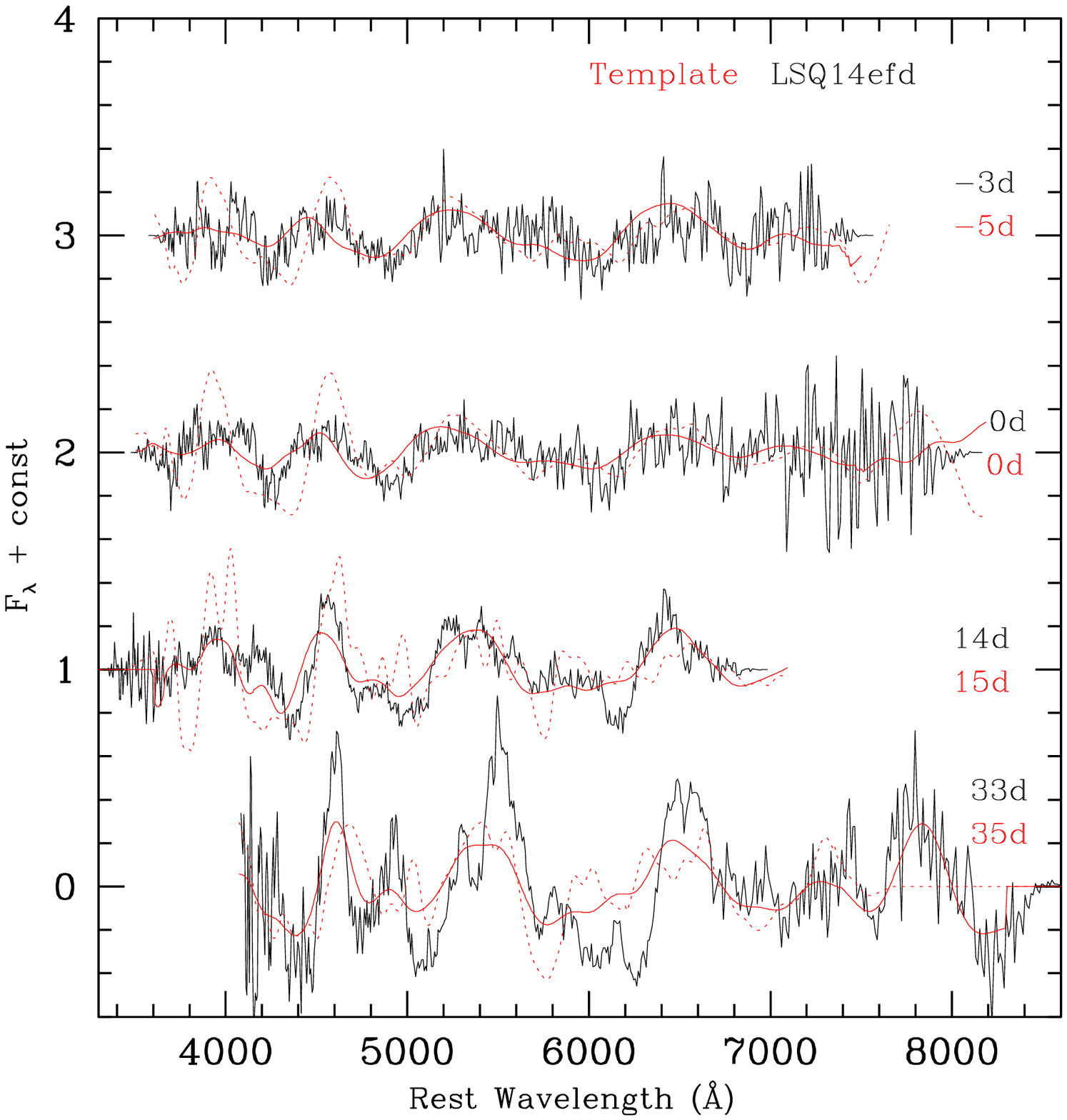}
	\end{center}
	\caption{Comparison of the spectra of LSQ14efd at different epochs with the templates \citep{Modjaz2014,Liu2016} for SNe-Ic (dashed red line) and SNe-Ic-BL (red solid line). The epochs refers to the V maximum. Spectra have been flattened through the SNID package \citep{Blondin2007} for the comparison with the templates.}
	\label{SpecTempl}
\end{figure*}

The ejecta velocities produced by the two different physical explosion mechanisms can be different. For comparison, ejecta velocities can be estimated  from a gaussian fit of the absorption profile of the P-Cygni features (after correction for the  redshift of the host galaxy). 
The uncertainties on the estimated velocities are a result of the error propagation on the uncertainties obtained from the measurement. This was confirmed by several repeated tests.
For LSQ14efd, the  $\sim -7$  spectrum indicates a Si II velocity of 
$v_{\rm ej} \sim 12300 \: \rm km \: s^{-1}$. At maximum, this fit to the same line gives   $v_{\rm ej} \sim 10000 \: \rm km \: s^{-1}$, which decreases to $ \sim 8000 \: \rm km \: s^{-1}$ after 10 days. At around $20$ days, we 
measure the Si II velocity  to drop to $ \sim 5000 \: \rm km \: s^{-1}$ and 
also estimate the Fe II lines to show an outflow of 
$\sim 3000 \: \rm km \: s^{-1}$ 
A comparison with SN 2004aw, shows that the Si II velocity, goes from $ \sim 12700 \: \rm km \: s^{-1}$ around maximum, to $ \sim 9300 \: \rm km \: s^{-1}$ after $10$ days, which are comparable with those estimated for of LSQ14efd,  within the errors. In the same temporal range, the Fe II velocity decreases from $\sim 12000 \: \rm km \: s^{-1}$ to $\sim 8100 \: \rm km \: s^{-1} $.
At around 20 day, SN 2004aw shows a Si II velocity of $ \sim 6000 \: \rm km \: s^{-1}$, in agreement with the value of Si II for LSQ14efd, within the errors.

\begin{figure}
  \begin{center}
  \includegraphics[scale=0.4]{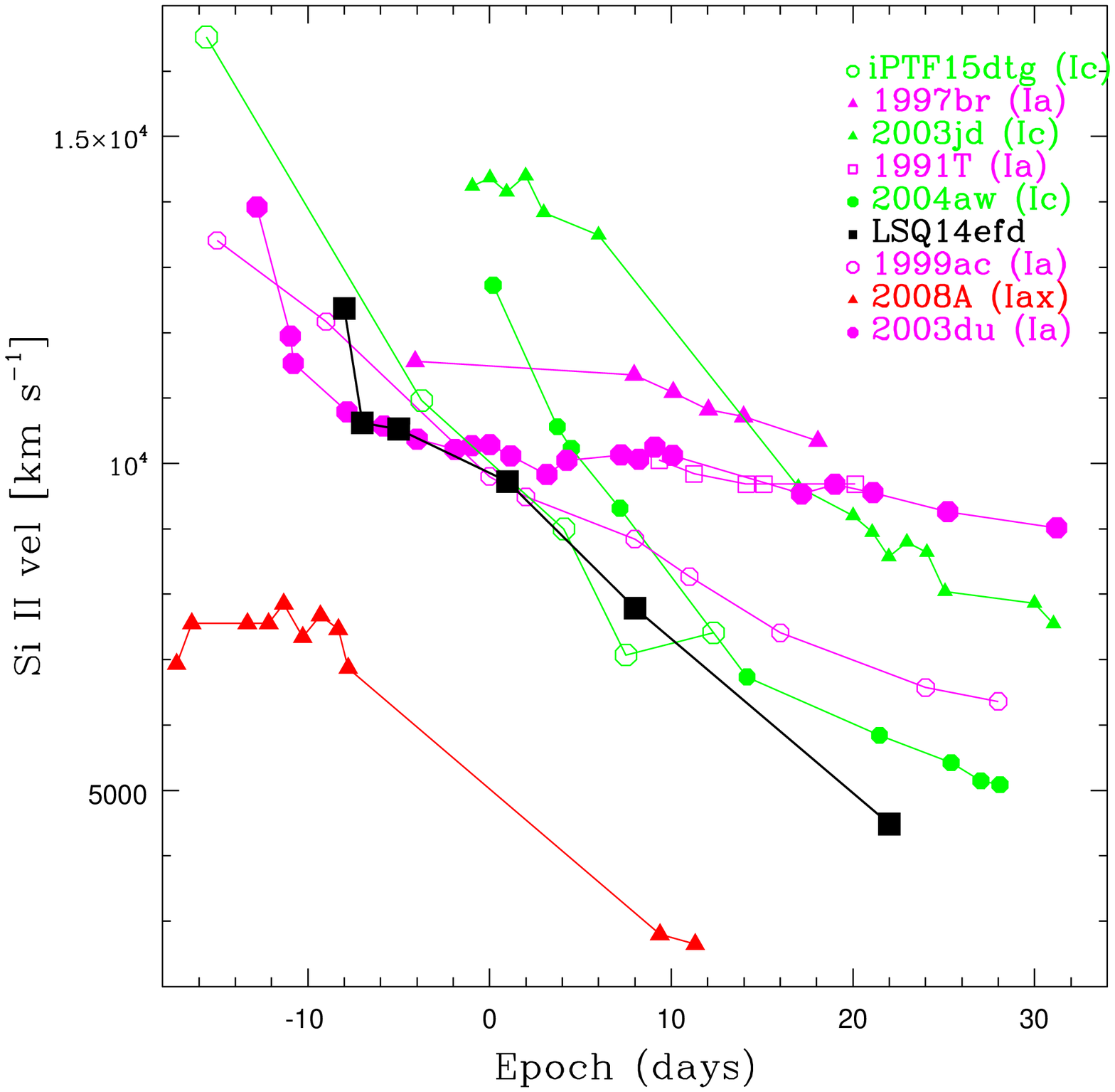}
  \end{center}
  \caption {Si II velocity evolution of LSQ14efd compared with type Ic SNe 2004aw \citep{Taubenberger2006}, 2003jd \citep{Valenti2008a}, iPTF15dtg \citep{Taddia2016} and with type Ia SNe 1999ac \citep{Garavini2005}, 1991T \citep{Phillips1992}, 1997br \citep{Li1999}, SN 2003du \citep{Stanishev2007} and type Iax 2008A \citep{Foley2013}. Epochs refer to th B-maximum.}
  \label{Si}
\end{figure}

\begin{figure}
	\centering
	\includegraphics[scale=0.43]{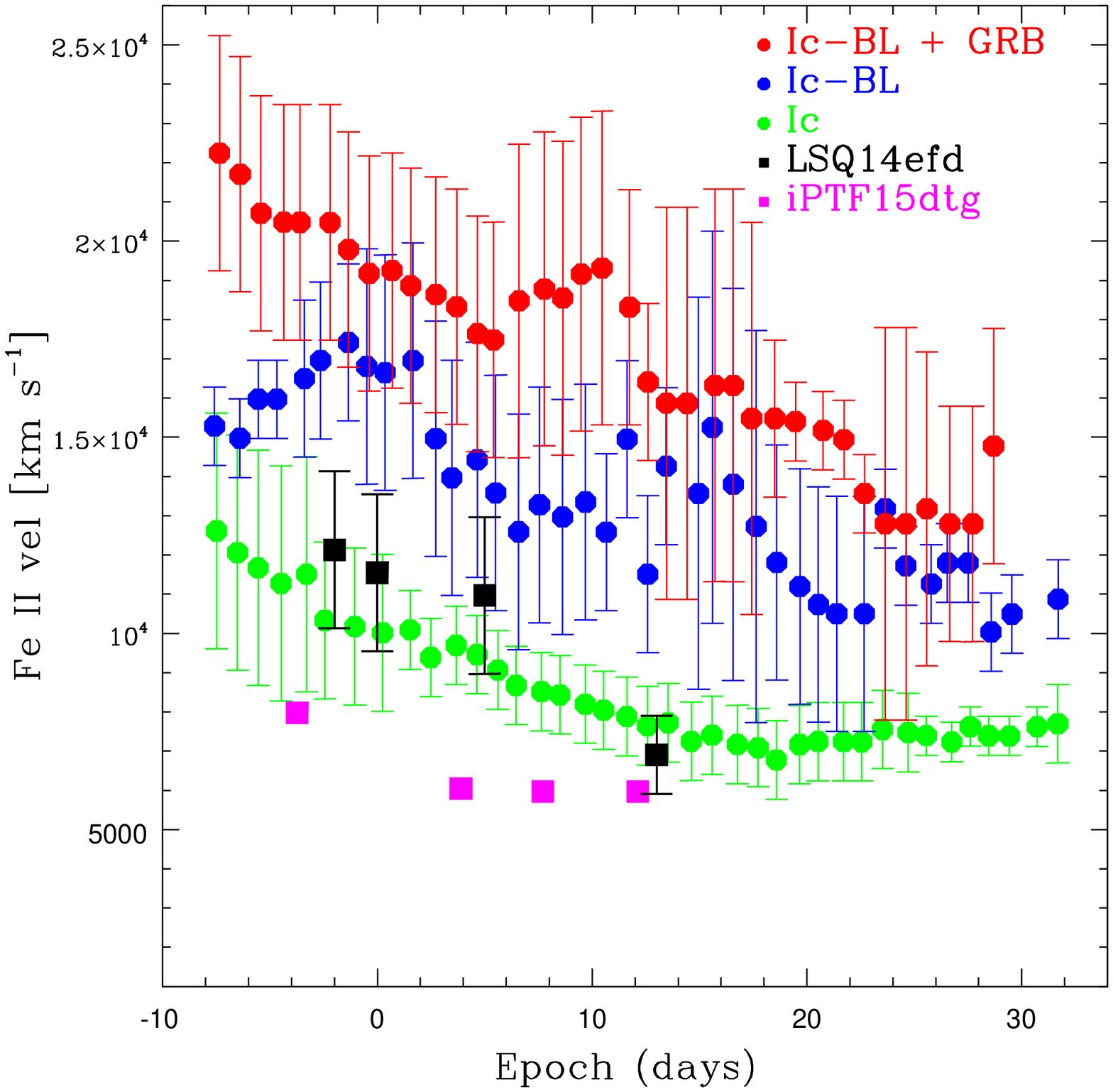}
	\caption{Comparison of Fe II $\lambda5169$ velocity of LSQ14efd with the general trend for SNe-Ic and SNe-Ic-BL found in \citealt{Modjaz2015}.
		Epochs refer to the V-maximum.}
	\label{FeII}
\end{figure}

We then compared the spectroscopic characteristics of LSQ14efd with those observed for sample of SNe-Ic \citep{Modjaz2015,Liu2016}.
In particular, the evolution of the Fe II (5169\AA) line velocities of LSQ14efd have been compared with the trends found by \citet{Modjaz2015} for SNe-Ic and SNe-Ic-BL, see Fig. \ref{FeII}.
We can notice that LSQ14efd shows an initial Fe II velocity slightly higher with respect to SNe-Ic trend but quite in agreement within the uncertainties.

Following the comparison done for the light curve of LSQ14efd with type Iax SNe we then proceed to study the Si II velocity evolution.

Fig. \ref{Si} shows the evolution of the SiII velocities of LSQ14efd compared with other type Ic SNe (SN 2004aw \citealt{Taubenberger2006}, SN 2003jd \citealt{Valenti2008a}, SN 1994I \citealt{Sauer2006}, SN 2007gr \citealt{Valenti2008b}, and SN 1998bw \citealt{Patat2001} other type Ia SNe 1999ac \citealt{Garavini2005}, 1991T \citealt{Phillips1992}, 1997br \citealt{Li1999} and SN 2003du \citealt{Stanishev2007}) and type Iax 2008A \citealt{Foley2013}). 

\citet{Benetti2005} found that in type Ia SNe there is a trend for  the Si II velocity to reach a value of $\sim 6000 \: {\rm km \: s}^{-1}$ at $\sim 30$ days.
Type Iax SNe, instead, are characterized by low-velocities of Si II. \cite{Foley2013} show, in their Fig. 19, the typical range of velocities for type Iax SNe to be $5000-8000 \: \rm km \: s^{-1}$, around maximum. Although 
 we can note that SN 2008A displays higher velocities than almost all the objects in this class ($\sim 8000 \rm \: km \: s^{-1}$).
Type Ic SNe, instead, show a wide diversity in ejecta velocities, varying from $\sim 30000 \: {\rm km \: s}^{-1}$ to $\sim 10000 \: {\rm km \:s}^{-1}$ around maximum. In summary, 
 LSQ14efd shows a velocity evolution (traced by Si II and Fe II)
which is  quantitatively similar to
that of  SN 2004aw. We find that  the velocities are compatible, within the errors, at almost all epoch. LSQ14efd velocities are significantly slower than those of the SN-Ic BL 2003jd at all epochs, but are faster than those of SN 2007gr. iPTF15dtg shows a higher pre-maximum velocity of Si II with respect to LSQ14efd, but starting from B-maximum they become comparable. The trend is also very similar to that of LSQ14efd.

The comparison with type Iax SN 2008A shows that the estimated Si II velocities for LSQ14efd have a similar slope but a higher value than the typical Iax SNe but they are low compared to that of type Ia SN.
We also note there is a clear difference in the slope of the Si II evolution between type Ia SNe and type Ic; the former have slower decrease. The slope of the velocity decrease for LSQ14efd is more  similar to those of type Ic SNe.

\section{$^{56}$N\lowercase{i} and ejected mass}

The light curve model developed by \citet{Arnett1982} gives an analytical description of light curve of type I SNe. The original paper by \citet{Arnett1982} contain a typographical error in the numerical factor of Arnett's equation 54, which is discussed in \citet{Lyman2016} and that has been corrected in equation \ref{ener}. The analysis performed at the peak of the light curve can lead to a rough estimate of the total mass of the SN ejecta ($M_{ej}$), the nickel mass ($M_{Ni}$) and the kinetic energy ($E_{kin}$).
The main assumptions of the model are: a homologous expansion; spherical symmetry; a constant optical opacity; no mixing of $^{56}Ni$ and radiation-pressure dominance. Furthemore, it considers also the diffusion approximation for photons, which can reasonably be applied in the early phases when the ejecta is optically thick, due to the high density.
The time-evolution of the SN luminosity is given by

\begin{equation}\label{nick}
L(t) = \epsilon_{Ni} \: e^{-x^{2}} \int\limits_{0}^{x} 2z \: e^{-2xy + z^{2}} dz
\end{equation}

\noindent where $x \equiv t/\tau_{m}$, $y \equiv \tau_{m}/2\tau_{Ni}$ and $\epsilon_{Ni} = Q_{Ni}/(M_{Ni} \: \tau_{Ni})$; $Q_{Ni}$ is the energy release for $^{56}Ni$ decay and $\tau_{Ni}$ is the $e-$folding time of the $^{56}Ni$ decay.
The width of the peak of the bolometric is related to the effective diffusion time and it is given by:

\begin{equation}\label{formula}
\tau_{m} = \left( \frac{2}{\beta c} \frac {k_{opt}M_{ej}}{v_{sc}} \right)^{1/2} \propto \: k_{opt}^{1/2} \: M_{ej}^{3/4} \: E_{kin}^{-1/4}
\end{equation}

\noindent where $v_{sc}$ is the velocity scale of the expansion, $k_{opt}$ is the optical opacity and $\beta$ is an integration constant.
Furthermore, assuming a homogeneous density of the ejecta, it is possible to relate the kinetic energy to the photospheric velocity ($v_{ph}$) at maximum through the relation \citep{Arnett1982}

\begin{equation}\label{ener}
v_{ph}^{2} \approx \frac{10}{3} \frac{E_{kin}}{M_{ej}}
\end{equation}

We first estimate the width of the light curve of LSQ14efd, $\tau \sim 20$ days, following the prescription in \citet{Lyman2016} so the estimate of the light curve width is the numbers of days required to reach the same magnitude the SN had 10 days prior maximum, then, assuming $ k_{opt} = 0.06 \: \rm{cm^{2} \: g^{-1}}$ \citep{Lyman2016} in equation \ref{formula} and considering the previously estimated Fe II velocity, we estimated $M_{ej} = 6.3 \pm 0.5 \: M_{\odot}$.

We then estimate the nickel mass synthesized through the equation \ref{nick}, evaluated at the time of the bolometric peak when we have 
reliable measurement. We estimate a ${M}_{Ni} = 0.25 \pm 0.06 \: M_{\odot}$. Finally, through equation \ref{ener}, we obtained an estimate for the kinetic energy, $E_{kin} = 5.6 \pm 0.5 \times 10^{51} \: \rm erg$. We note that since the ${M}_{Ni}$ depends on the peak luminosity and  has been estimated from a quasi-bolometric light curve, we should consider
this value  as a lower limit.
Considering the peak luminosity inferred from the method developed by \citet{Lyman2014}, the nickel mass could increase up to $\sim 0.32 \: M_{\odot}$.
The estimated physical parameters are generally higher than other SNe-Ic, excpet for the nickel mass which is close to that estimated for SNe 2004aw \citep{Taubenberger2006} and 2003jd \citep {Valenti2008a}. We note that a recent work from \citealt{Mazzali2017} revisited the explosion parameters for SN 2004aw but they are still in agreement, within the error, with the ones from \citealt{Taubenberger2006} and used in this work.
As expected, from the empirical comparison of the light curves, the 
the ejecta mass and $^{56}$Ni mass are 
larger than those found for type Ic SN 1994I \citep{Nomoto1994}, and somewhat comparable with the higher value for broad-lined type Ic such  as SN 1998bw (\citealt{Galama1998}, \citealt{Nakamura2000}). iPTF15dtg is more energetic and has a more massive evelope and nickel mass than LSQ14efd. Table \ref{param} contains a quantitative comparison.
A comparison of the explosion paramters with the average value found by \citet{Lyman2016} is also shown in Table \ref{param}. We note how LSQ14efd has generally higher value than the average found for normal SNe-Ic but somewhat similar to that found for SNe-Ic-BL. We also note that the LSQ14efd velocity is higher than the average value found for SNe-Ic but still lower than those of SNe-Ic-BL.
We point out that the kinetic energy estimated has to be considered as an upper limit since spherical symmetry was adopted for the explosion model, for all these SNe in Table \ref{param} . Instead, \citet{Taubenberger2009} show that more than half of all stripped-envelope CC-SNe explosion may be significantly aspherical.

\begin{table}
	\caption{Comparison of the physical parameters of some type Ic SNe and average values from \citet{Lyman2016}: photospheric velocity, nickel mass, mass of the ejecta and kinetic energy.\label{param}}
	\setlength{\tabcolsep}{0.5pt}.
	\begin{footnotesize}
		\centering
		\begin{tabular}{lccccc}
			\hline
		SN & $M_{R}$ & $v_{ph}^{a}$ &  $M_{Ni}$ & $M _{ej}$ & $E_{kin}$ \\
		& mag &  $10^{3} \rm \: km \: s^{-1}$ &  $M_{\odot}$ & $M_{\odot}$ & $10^{51} \: \rm erg$ \\
		\hline
		1994I & -17.7 & 11 &  0.07 & 0.9 & 1 \\
		2007gr & -17.3 & 11 & 0.07-0.1 & 1.5-3 & 1.5-3 \\
		2004aw & -18.14 & 14 & 0.25-0.35 & 3.5 - 8.0 & 3.5 - 9.0 \\
		iPTF15dtg & -18.51 & 6 & 0.41-0.43 & 9.7-10.9 & 2.1-2.3 \\
		LSQ14efd  & -18.6 & 12.2 & 0.25 & 6.3 & 5.3 \\
		2003jd & -19 & 13 & 0.36 & 3.0 & 5 - 10 \\
		1998bw & -19.14 & 18 & 0.7 & 10 & 30 \\
		Average Ic & & 10.4 & 0.22 & 3.0 & 1.9 \\
		Average Ic-BL & & 19.1 & 0.32 & 2.9 & 6.0 \\
		\hline
			$^{a}$ At the B-maximum 
		\end{tabular}
		\\[1.5ex]
		
	\end{footnotesize}
\end{table}

\begin{figure}
	\centering
	\includegraphics[scale=0.4]{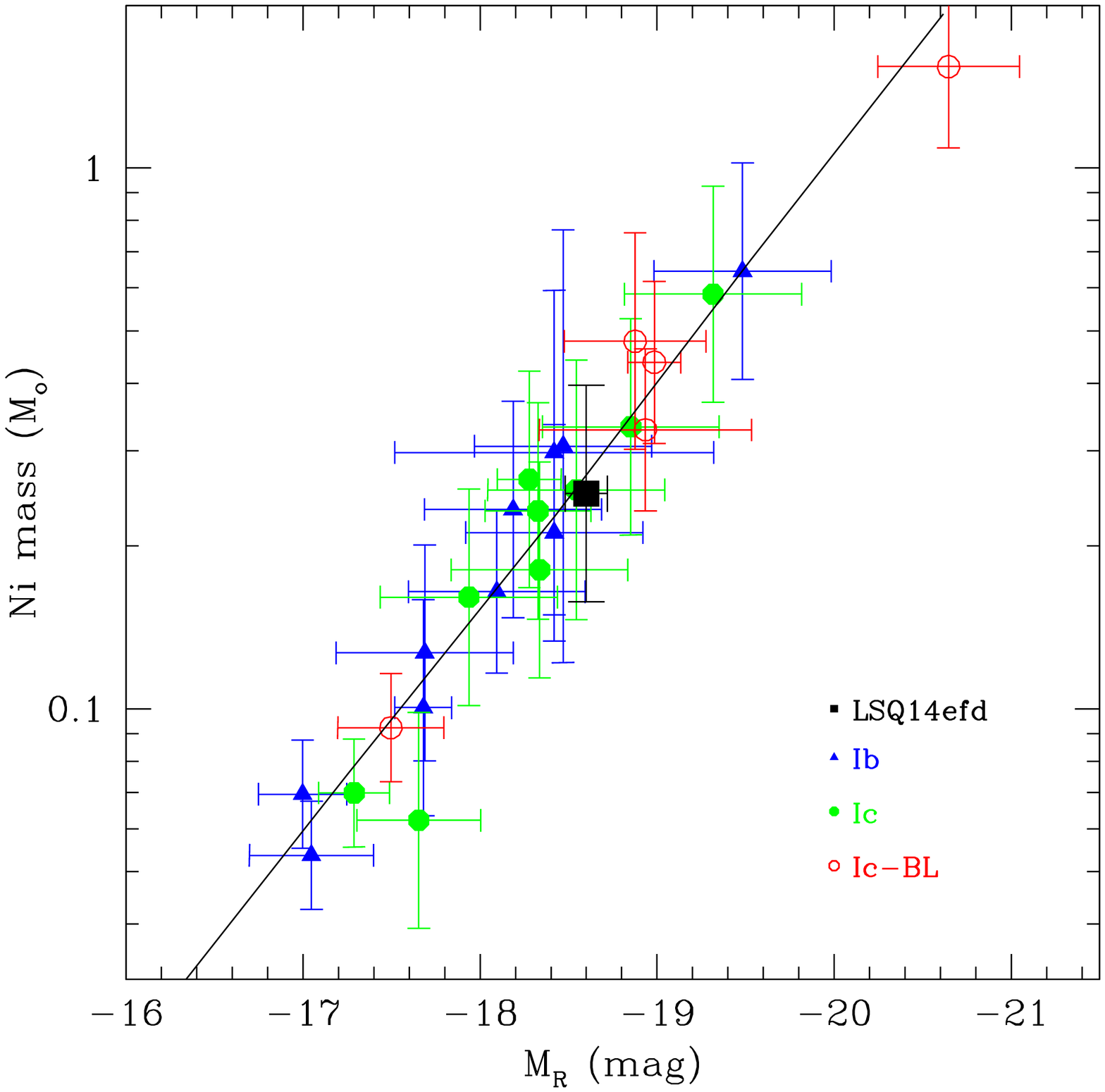}
	\caption{Comparison of Ni mass and $M_{R}$ of LSQ14efd with the general trend for SNe-Ib, Ic and Ic-BL found in \citealt{Drout2011}. the solid line shows the best fit.
	}
	\label{M_ni}
\end{figure}

As a consistency check, we have also estimated the explosion parameters assuming that LSQ14efd is a possible type Ia SN using the same model but different assumptions for the physical constants which are more 
appropriate for an exploding CO white dwarf.
In this case we considered $k_{opt} = 0.3 \rm \; g \; cm^{-2} \; s^{-2}$ \citep{Stritzinger2006} and therefore obtained a different ejecta mass
of $M_{ej} = 1.0 \pm 0.2 \: M_{\odot}$ and $E_{kin} = 0.6 \pm 0.2 \times 10^{51} \: erg$. The estimated nickel mass  still remains at $0.25 \pm 0.06 \: M_{\odot}$. As expected, the  difference in the values of the $M_{ej}$ and $E_{kin}$ under the two assumptions is simply due to the different values of $k_{opt}$, while the $M_{Ni}$ remains the same as it depends on the peak luminosity.
We compared these results also with those of type Iax SN as reported in \citealt{Magee2017}, where the ${M}_{Ni}$ for SNe-Iax ranges in $\sim 0.03 - 0.6 \: M_{\odot}$. The estimated ${M}_{Ni}$ for LSQ14efd falls in this interval and it's not possible to use this parameter to discriminate among the two possible scenarios.
We then compared the results obtained for LSQ14efd with those of a sample of SNe-Ic \citep{Drout2011}. In particular, the Ni mass with respect to the absolute magnitude in R band $M_{R}$ of LSQ14efd was compared with the trend found by \citet{Drout2011} for SNe-Ic and SNe-Ic-BL (Fig. \ref{M_ni}). \citet{Drout2011} derived the $^{56}Ni$ mass through the light curve models from \citet{Valenti2008a} which are based on \citet{Arnett1982} formalism. We can notice that LSQ14efd follows the trend and it is in agreement with the best fit evolution, within the uncertainties.

\section{Conclusion}\label{discussion}

We have presented the photometric and spectroscopic follow-up of  LSQ14efd within the PESSTO survey, which covered a period of $\sim 100$ days. LSQ14efd exploded in an anonymous galaxy at a distance modulus $\mu= 37.35 \pm 0.15$ mag and  does not  appear to suffer of strong reddening ($E(B-V)= 0.0376 \pm 0.0015$ mag). \\
Early photometric observations show the probable detection of the cooling of the shock break-out event in the light curve of LSQ14efd. A comparison with other CC-SNe shows a similarity in the cooling of the shock break-out detection of LSQ14efd and SN 2008D (see Fig.\,\ref{shock}).
The well sampled colour evolution was studied to investigate the possibility that LSQ14efd is a SN-Ia with an indication of interaction of its ejecta with either a companion or nearby CSM, such as as SN 2012cg and iPTF14atg (see Section\,\ref{photoanalysis}). \\
We want to stress that this early emission can have also different interpretation for CC-SNe as the presence of an extended envelope previously ejected by the progenitor or it can be due to some outwardly mixed $^{56}Ni$. But with only a single data point and in a single band it is actually quite difficult to definitely distinguish the source of this early emission and then discriminate among different possible progenitor scenarios, either CC-SNe or SNe-Iax.

We presented an analysis of the main photometric features of this SN 
including multicolor light curves, colour evolution and bolometric light curves, (see Figs. \ref{shock} (right panel), \ref{color_all} and \ref{bolom_all}).

We point out that 
a characteristic property of the photometric evolution of  LSQ14efd is represented by the shift in time between the peak in the red and blue bands with an offset of $\sim 8.9$ days between B and I bands, very likely due to different time scales in the cooling of the ejecta. Another characteristic is depicted by the slow decline rate of the light curves (in the interval 5-30 days post B-maximum, $5.61 \pm 0.78$ mag per 100 days in the B band and $3.28 \pm 0.33$ mag per 100 days in the R band), see Table \ref{comp2004aw}. \\
The colour evolutions of LSQ14efd (Fig. \ref{color_all}) resemble those of type Ic SNe but do not differ much from SNe Iax as well. The quasi-bolometric light curve (Fig. \ref{bolom_all}) is similar to those of type Ic SNe 2004aw and 2003jd but also comparable with type Iax SN 2008A. \\
We also perform a comparison of some observables of LSQ14efd with the trend found for sample of SNe-Ic \citep{Drout2011,Modjaz2015,Liu2016}. In particular LSQ14efd R band follows the template evolution found in \citet{Drout2011} while the V band instead differs significantly in the pre maximum evolution, showing that LSQ14efd has a broader light curve than average SN-Ib/c (Fig. \ref{R_template}). LSQ14efd follow the  $^{56}Ni$ mass vs $M_{R}$ correlation found by \citet{Drout2011}. \\
The spectroscopic evolution of LSQ14efd shows Si II, Fe II lines and a likely Ca II feature and it is similar to that of SN 2004aw at epochs that
are consistent with the light curve evolution. 
The evolution of LSQ14efd Si II velocities (see Fig. \ref{Si}) from B-maximum ($\sim 10000 \: \rm km s^{-1}$) until 20 days after ($5000 \: \rm km \: s^{-1}$) is very similar to that of SN 2004aw, which is intermediate between ``standard'' SNe-Ic and the very energetic ones, like SN 1998bw. Recently, it has also been proposed that SN 2004aw is a ``fast-lined" SN rather than a BL-SN \citep{Mazzali2017}, strengthening the peculiarity of this SN.
However the spectra of this object present some similarities with peculiar type Ia and Iax SNe, in particular in the late phase (see Figs. \ref{comp1999ac_08A} and \ref{comp_all}) and based on spectra only, we can not determine wheter this is a peculiar SN-Ia or a Ic.
A comparison with SNe-Ic and SNe-Ic-BL templates \citep{Liu2016,Modjaz2014} remarks a quite good similarity with SNe-Ic for the early spectra but a significant deviation for the late ones but a better agreement with the templates for SNe-Ic-BL at all epochs. This strengthens the peculiarity of LSQ14efd. \\
The Fe II evolution of LSQ14efd was compared with the average trend for SNe-Ic found by \citealt{Modjaz2015} for SNe-Ic and it shows a comparable behaviour at early epochs.
Instead LSQ14efd shows a pretty different Si II evolution with respect of SNe-Ia and Iax. \\
Considering the overall proprierties shown by LSQ14efd we favour a core-collapse origin for LSQ14efd.

We applied a simple model for CC-SNe to the quasi-bolometric light curve to calculate the physical parameters of LSQ14efd. We obtained a synthesized ${M}_{Ni} = 0.25 \pm 0.06 \: M_{\odot}$, an ejected mass of $6.3 \: M_{\odot}$ and a kinetic energy of $5.6 \times 10^{51} \: erg$.
A comparison of the explosion parameters with the average values found by \citet{Lyman2016} shows that LSQ14efd seems to have values closer to that found for SNe-Ic-BL rather than those for normal SNe-Ic.
No evident association with a GRB was identified.

The increasing number of discoveries of peculiar SNe-Ic, which represent a link between energetic Ic events which are not connected with GRBs (as SN 2003jd) and those which show a clear association with GRBs (as GRB 980425/SN 1998bw) gives support to the idea of an existing continuum of properties between  broad lined Ic and ``standard'' SNe-Ic events rather than suggesting the existence of clearly ``separated'' classes of SNe-Ic.
LSQ14efd is more energetic than standard SNe-Ic, it is not as energetic as SNe-Ic-BL which are often associated with GRBs. This again shows a diversity of the type Ic SNe class, which was already pointed out by \citet{Taubenberger2006}.

LSQ14efd confirms the existence of an unresolved ambiguity in SN classification, particularly when the classification in SN types relies only on the photometric evolution and/or early stages spectra.

 \section{Acknowledgements}

We warmly thank our referee for the helpful comments which significanly improved the content and the readability of our manuscript.
We thank S. Taubenberger and J. Parrent for the useful discussion and comments.
C.B. thanks the IRAP PhD program for the financial support.
GP acknowledge support provided by the Millennium Institute of Astrophysics (MAS) through grant IC120009 of the Programa Iniciativa Cientifica Milenio del Ministerio de Economia, Fomento y Turismo de Chile".
M.D.V., M.L.P., S.B., A.P., L.T. and M.T. are  partially supported by the PRIN-INAF 2014 with the project "Transient Universe: unveiling new types of stellar explosions with PESSTO.
This work was partly supported by the European Union FP7 programme through ERC grant number 320360.
This material is based upon work supported by the National Science Foundation under grant number 1313484.

This work is based (in part) on observations collected at the European Organisation for Astronomical Research in the Southern Hemisphere, Chile as part of PESSTO, (the Public ESO Spectroscopic Survey for Transient Objects Survey) ESO program 188.D-3003, 191.D-0935.
The research leading to these results has received funding from the European Research Council under the European Union's Seventh Framework Programme (FP7/2007-2013)/ERC Grant agreement n$^{\rm o}$ [291222]  (PI : S. J. Smartt) and STFC grants ST/I001123/1 and ST/L000709/1.
This paper is partially based on observations collected obtained through the CNTAC proposal CN2014A-101 and CN2014B-81.


\begin{thebibliography}{}

\bibitem[Arcavi et al.(2011)]{Arcavi2011}
Arcavi, I., Gal-Yam, A., Yaron, O., et al.\ 2011, \apjl, 742, L18

\bibitem[Arnett(1982)]{Arnett1982}
Arnett, W.~D.\ 1982, \apj, 253, 785

\bibitem[Arnett et al.(1989)]{Arnett1989}
Arnett, W.~D., Bahcall, J.~N., Kirshner, R.~P., \& Woosley, S.~E.\ 1989, \araa, 27, 629

\bibitem[Baltay et al.(2012)]{Baltay2012} Baltay, C., Rabinowitz, 
D., Hadjiyska, E., et al.\ 2012, The Messenger, 150, 34 

\bibitem[Baltay et al.(2013)]{Baltay2013}
Baltay, C., Rabinowitz, D., Hadjiyska, E., et al.\ 2013, \pasp, 125, 683

\bibitem[Benetti et al.(2005)]{Benetti2005}
Benetti, S., Cappellaro, E., Mazzali, P.~A., et al.\ 2005, \apj, 623, 1011

\bibitem[Blondin et al.(2012)]{Blondin2012}
Blondin, S., Matheson, T., Kirshner, R.~P., et al.\ 2012, \aj, 143, 126

\bibitem[Blondin \& Tonry(2007)]{Blondin2007} Blondin, S., \& Tonry, J.~L.\ 2007, \apj, 666, 1024

\bibitem[Bersten et al.(2014)]{Bersten2014}
Bersten, M.~C., Benvenuto, O.~G., Folatelli, G., et al.\ 2014, \aj, 148, 68

\bibitem[Branch et al.(2002)]{Branch2002}
Branch, D., Benetti, S., Kasen, D., et al.\ 2002, \apj, 566, 1005

\bibitem[Campana et al.(2006)]{Campana2006}
Campana, S., Mangano, V., Blustin, A.~J., et al.\ 2006, \nat, 442, 1008

\bibitem[Cano et al.(2014)]{Cano2014}
Cano, Z., de Ugarte Postigo, A., Pozanenko, A., et al.\ 2014, \aap, 568, A19

\bibitem[Cao et al.(2013)]{Cao2013}
Cao, Y., Kasliwal, M.~M., Arcavi, I., et al.\ 2013, \apjl, 775, L7

\bibitem[Cao et al.(2015)]{Cao2015}
Cao, Y., Kulkarni, S.~R., Howell, D.~A., et al.\ 2015, \nat, 521, 328

\bibitem[Cappellaro et al.(2015)]{Cappellaro2015}
Cappellaro, E., Botticella, M.~T., Pignata, G., et al.\ 2015, arXiv:1509.04496 
 
\bibitem[Cardelli et al.(1989)]{Cardelli1989}
Cardelli, J.~A., Clayton, G.~C., \& Mathis, J.~S.\ 1989, \apj, 345, 245

\bibitem[Della Valle(2011)]{MDV2011}
Della Valle, M.\ 2011, International Journal of Modern Physics D, 20, 1745

\bibitem[Drout et al.(2016)]{Drout2016} Drout, M.~R., Milisavljevic, D., Parrent, J., et al.\ 2016, \apj, 821, 57 

\bibitem[Drout et al.(2011)]{Drout2011} Drout, M.~R., Soderberg, A.M., Gal-Yam, A., et al.\ 2011, \apj, 741, 97

\bibitem[Eldridge et al.(2013)]{Eldridge2013}
Eldridge, J.~J., Fraser, M., Smartt, S.~J., Maund, J.~R., \& Crockett, R.~M.\ 2013, \mnras, 436, 774

\bibitem[Eldridge et al.(2015)]{Eldridge2015} Eldridge, J.~J., Fraser, M., Maund, J.~R., \& Smartt, S.~J.\ 2015, \mnras, 446, 2689

\bibitem[Elmhamdi et al.(2006)]{Elmhamdi2006}
Elmhamdi, A., Danziger, I.~J., Branch, D., et al.\ 2006, \aap, 450, 305
  
\bibitem[Filippenko(1997)]{Filippenko1997}
Filippenko, A.V.\ 1997, ARAA, 35, 309

\bibitem[Foley et al.(2009)]{Foley2009}
Foley, R.~J., Chornock, R., Filippenko, A.~V., et al.\ 2009, \aj, 138, 376

\bibitem[Foley et al.(2013)]{Foley2013}
Foley, R.~J., Challis, P.~J., Chornock, R., et al.\ 2013, \apj, 767, 57

\bibitem[Fremling et al.(2014)]{Fremling2014}
Fremling, C., Sollerman, J., Taddia, F., et al.\ 2014, \aap, 565, A114
  
\bibitem[Galama et al.(1998)]{Galama1998}
Galama, T.~J., Vreeswijk, P.~M., van Paradijs, J., et al.\ 1998, \nat, 395, 670
  
\bibitem[Garavini et al.(2005)]{Garavini2005}
Garavini, G., Aldering, G., Amadon, A., et al.\ 2005, \aj, 130, 2278

\bibitem[Groh et al.(2013)]{Groh2013}
Groh, J.~H., Georgy, C., \& Ekstr{\"o}m, S.\ 2013, \aap, 558, L1 

\bibitem[Guetta \& Della Valle (2007)]{guetta07}
Guetta, D. \& Della Valle 2007, ApJ, 657, L73

\bibitem[Hillebrandt \& Niemeyer(2000)]{Hillebrandt2000}
Hillebrandt, W., \& Niemeyer, J.~C.\ 2000, \araa, 38, 191

\bibitem[Howell et al.(2006)]{Howell2006}
Howell, D.~A., Sullivan, M., Nugent, P.~E., et al.\ 2006, \nat, 443, 308

\bibitem[Hunter et al.(2009)]{Hunter2009} Hunter, D.~J., Valenti, S., Kotak, R., et al.\ 2009, \aap, 508, 371

\bibitem[Iben \& Tutukov(1984)]{Iben1984}
Iben, I., Jr., \& Tutukov, A.~V.\ 1984, \apjs, 54, 335

\bibitem[Kasen(2010)]{Kasen2010}
Kasen, D.\ 2010, \apj, 708, 1025

\bibitem[Kasliwal et al.(2010)]{Kasliwal2010}
Kasliwal, M.~M., Kulkarni, S.~R., Gal-Yam, A., et al.\ 2010, \apjl, 723, L98

\bibitem[Katz \& Dong(2012)]{Katz2012}
Katz, B., \& Dong, S.\ 2012, arXiv:1211.4584
 
\bibitem[Kovacevic et al.(2014)]{Kovacevic2014}
Kovacevic, M., Izzo, L., Wang, Y., et al.\ 2014, \aap, 569, A108

\bibitem[Labbe et al.(2001)]{Labbe2001}
Labbe, E., Galaz, G., Krisciunas, K., et al.\ 2001, Bulletin of the American Astronomical 
  Society, 33, 1370 
  
\bibitem[Leaman et al.(2011)]{Leaman2011}
Leaman, J., Li, W., Chornock, R., \& Filippenko, A.~V.\ 2011, \mnras, 412, 1419 

\bibitem[Lewis et al.(1994)]{Lewis1994}
Lewis, J.~R., Walton, N.~A., Meikle, W.~P.~S., et al.\ 1994, \mnras, 266, L27

\bibitem[Li et al.(1999)]{Li1999}
Li, W.~D., Qiu, Y.~L., Qiao, Q.~Y., et al.\ 1999, \aj, 117, 2709

\bibitem[Li et al.(2001)]{Li2001}
Li, W., Filippenko, A.~V., \& Riess, A.~G.\ 2001, \apj, 546, 719

\bibitem[Li et al.(2003)]{Li2003} Li, W., Filippenko, A.~V., Chornock, R., et al.\ 2003, \pasp, 115, 453 

\bibitem[Li et al.(2011)]{Li2011}
Li, W., Leaman, J., Chornock, R., et al.\ 2011, \mnras, 412, 1441

\bibitem[Liu et al.(2016)]{Liu2016} Liu, Y.-Q., Modjaz, M., Bianco, F.~B., \& Graur, O.\ 2016, \apj, 827, 90
 
\bibitem[Lupton et al.(2005)]{Lupton2005}
Lupton, R.~H., Juri{\'c}, M., Ivezi{\'c}, Z., et al.\ 2005, Bulletin of the American 
 Astronomical Society, 37, \#133.08
 
\bibitem[Lyman et al.(2014)]{Lyman2014}
Lyman, J.~D., Bersier, D., \& James, P.~A.\ 2014, \mnras, 437, 3848 

\bibitem[Lyman et al.(2016)]{Lyman2016} Lyman, J.~D., Bersier, D., James, P.~A., et al.\ 2016, \mnras, 457, 328

\bibitem[Magee et al.(2017)]{Magee2017} Magee, M.~R., Kotak, R., Sim, S.~A., et al.\ 2017, arXiv:1701.05459 

\bibitem[Marion et al.(2015)]{Marion2015}
Marion, G.~H., Brown, P.~J., Vink{\'o}, J., et al.\ 2015, arXiv:1507.07261

\bibitem[Mazzali et al.(2017)]{Mazzali2017} Mazzali, P.~A., Sauer, D.~N., Pian, E., et al.\ 2017, \mnras, 469, 2498

\bibitem[Mazzali et al.(2008)]{Mazzali2008}
Mazzali, P.~A., Valenti, S., Della Valle, M., et al.\ 2008, Science, 321, 1185
 
\bibitem[Mannucci et al.(2006)]{Mannucci2006}
Mannucci, F., Della Valle, M., \& Panagia, N.\ 2006, \mnras, 370, 773
 
\bibitem[McCully et al.(2014)]{McCully2014}
McCully, C., Jha, S.~W., Foley, R.~J., et al.\ 2014, \apj, 786, 134
 
\bibitem[Modjaz et al.(2015)]{Modjaz2015} Modjaz, M., Liu, Y.~Q., Bianco, F.~B., \& Graur, O.\ 2015, arXiv:1509.07124 

\bibitem[Modjaz et al.(2014)]{Modjaz2014} Modjaz, M., Blondin, S., Kirshner, R.~P., et al.\ 2014, \aj, 147, 99
 
\bibitem[Munari et al.(2014)]{Munari2014}
Munari, U., Henden, A., Frigo, A., et al.\ 2014, \aj, 148, 81

\bibitem[Nakamura et al.(2000)]{Nakamura2000}
Nakamura, T., Nomoto, K., Iwamoto, K., et al.\ 2000, MmSAI, 71, 345

\bibitem[Nakamura et al.(2001)]{Nakamura2001}
Nakamura, T., Mazzali, P.~A., Nomoto, K., \& Iwamoto, K.\ 2001, \apj, 550, 991

\bibitem[Nakar \& Sari(2012)]{Nakar2012} Nakar, E., \& Sari, R.\ 2012, \apj, 747, 88

\bibitem[Nomoto et al.(1994)]{Nomoto1994}
Nomoto, K., Yamaoka, H., Pols, O.~R., et al.\ 1994, \nat, 371, 227

\bibitem[Pakmor et al.(2008)]{Pakmor2008}
Pakmor, R., R{\"o}pke, F.~K., Weiss, A., \& Hillebrandt, W.\ 2008, \aap, 489, 943

\bibitem[Panagia \& Laidler(1991)]{Panagia1991}
Panagia, N., \& Laidler, V.~G.\ 1991, Wolf-Rayet Stars and Interrelations with Other Massive Stars in Galaxies, 143, 567
  
\bibitem[Parrent et al.(2015)]{Parrent2015}
Parrent, J.~T., Milisavljevic, D., Soderberg, A.~M., \& Parthasarathy, M.\ 2015, arXiv:1505.06645
  
\bibitem[Patat et al.(2001)]{Patat2001}
Patat, F., Cappellaro, E., Danziger, J., et al.\ 2001, \apj, 555, 900

\bibitem[Perets et al.(2010)]{Perets2010}
Perets, H.~B., Gal-Yam, A., Mazzali, P.~A., et al.\ 2010, \nat, 465, 322

\bibitem[Phillips et al.(1992)]{Phillips1992}
Phillips, M.~M., Wells, L.~A., Suntzeff, N.~B., et al.\ 1992, \aj, 103, 1632 

\bibitem[Phillips(1993)]{Phillips1993}
Phillips, M.~M.\ 1993, \apjl, 413, L105
 
\bibitem[Phillips et al.(2006)]{Phillips2006}
Phillips, M.~M., Krisciunas, K., Suntzeff, N.~B., et al.\ 2006, \aj, 131, 2615

\bibitem[Phillips et al.(2007)]{Phillips2007}
Phillips, M.~M., Li, W., Frieman, J.~A., et al.\ 2007, \pasp, 119, 360

\bibitem[Poznanski(2010)]{Poznanski2010}
Poznanski, D.\ 2010, Progenitors and Environments of Stellar Explosions, 50

\bibitem[Prentice et al.(2016)]{Prentice2016} Prentice, S.~J., Mazzali, P.~A., Pian, E., et al.\ 2016, \mnras, 

\bibitem[Rabinak et al.(2012)]{Rabinak2012} Rabinak, I., Livne, E., \& Waxman, E.\ 2012, \apj, 757, 35

\bibitem[Rajala et al.(2004)]{Rajala2004}
Rajala, A., Fox, D.~B., \& Gal-Yam, A.\ 2004, The Astronomer's Telegram, 320, 1

\bibitem[Rajala et al.(2005)]{Rajala2005}
Rajala, A.~M., Fox, D.~B., Gal-Yam, A., et al.\ 2005, \pasp, 117, 132
  
\bibitem[Richmond et al.(1996)]{Richmond1996}
Richmond, M.~W., van Dyk, S.~D., Ho, W., et al.\ 1996, \aj, 111, 327
  
\bibitem[Sauer et al.(2006)]{Sauer2006}
Sauer, D.~N., Mazzali, P.~A., Deng, J., et al.\ 2006, \mnras, 369, 1939

\bibitem[Scalzo et al.(2010)]{Scalzo2010}
Scalzo, R.~A., Aldering, G., Antilogus, P., et al.\ 2010, \apj, 713, 1073
 
\bibitem[Schlegel et al.(1998)]{Schlegel1998}
Schlegel, D.~J., Finkbeiner, D.~P., \& Davis, M.\ 1998, \apj, 500, 525 

\bibitem[Serduke et al.(2006)]{Serduke2006}
Serduke, F.~J.~D., Blum, A., Scala, J., Filippenko, A.~V., \& Chornock, R.\ 2006, Central Bureau Electronic Telegrams, 380, 1

\bibitem[Shelton(1993)]{Shelton1993} Shelton, I.~K.\ 1993, \aj, 105, 1895 

\bibitem[Silverman et al.(2011)]{Silverman2011}
Silverman, J.~M., Ganeshalingam, M., Li, W., et al.\ 2011, \mnras, 410, 585

\bibitem[Silverman et al.(2012)]{Silverman2012}
Silverman, J.~M., Ganeshalingam, M., Cenko, S.~B., et al.\ 2012, \apjl, 756, L7

\bibitem[Soderberg et al.(2008)]{Soderberg2008} Soderberg, A.~M., Berger, E., Page, K.~L., et al.\ 2008, \nat, 453, 469 
 
\bibitem[Smartt(2009)]{Smartt2009}
Smartt, S.~J.\ 2009, \araa, 47, 63
 
\bibitem[Smartt et al.(2015)]{Smartt2015}
Smartt, S.~J., Valenti, S., Fraser, M., et al.\ 2015, \aap, 579, A40

\bibitem[Smartt(2015)]{Smartt2015b}
Smartt, S.~J.\ 2015, PASA, 32, e016

\bibitem[Stanishev et al.(2007)]{Stanishev2007}
Stanishev, V., Goobar, A., Benetti, S., et al.\ 2007, \aap, 469, 645
 
\bibitem[Stetson(1987)]{Stetson1987}
Stetson, P.~B.\ 1987, \pasp, 99, 191

\bibitem[Stritzinger et al.(2006)]{Stritzinger2006}
Stritzinger, M., Leibundgut, B., Walch, S., \& Contardo, G.\ 2006, \aap, 450, 241

\bibitem[Sullivan et al.(2011)]{Sullivan2011}
Sullivan, M., Kasliwal, M.~M., Nugent, P.~E., et al.\ 2011, \apj, 732, 118

\bibitem[Taddia et al.(2016)]{Taddia2016} Taddia, F., Fremling, C., Sollerman, J., et al.\ 2016, arXiv:1605.02491

\bibitem[Tartaglia et al.(2014a)]{Tartaglia2014a}
Tartaglia, L., Elias-Rosa, N., Morales-Garoffolo, A., et al.\ 2014, The Astronomer's 
Telegram, 6400, 1

\bibitem[Tartaglia et al.(2014b)]{Tartaglia2014b}
Tartaglia, L., Elias-Rosa, N., Morales-Garoffolo, A., et al.\ 2014, The Astronomer's 
Telegram, 6403, 1 

\bibitem[Taubenberger et al.(2006)]{Taubenberger2006}
Taubenberger, S., Pastorello, A., Mazzali, P.~A., et al.\ 2006, \mnras, 371, 1459

\bibitem[Taubenberger et al.(2009)]{Taubenberger2009}
Taubenberger, S., Valenti, S., Benetti, S., et al.\ 2009, \mnras, 397, 677

\bibitem[Taubenberger et al.(2011)]{Taubenberger2011}
Taubenberger, S., Benetti, S., Childress, M., et al.\ 2011, \mnras, 412, 2735

\bibitem[Turatto et al.(1996)]{Turatto1996}
Turatto, M., Benetti, S., Cappellaro, E., et al.\ 1996, \mnras, 283, 1

\bibitem[Tutukov \& Yungelson(1979)]{Tutukov1979}
Tutukov, A.~V., \& Yungelson, L.~R.\ 1979, \actaa, 29, 665

\bibitem[Valenti et al.(2014)]{Valenti2014}
Valenti, S., Yuan, F., Taubenberger, S., et al.\ 2014, \mnras, 437, 1519

\bibitem[Valenti et al.(2011)]{Valenti2011}
Valenti, S., Fraser, M., Benetti, S., et al.\ 2011, \mnras, 416, 3138

\bibitem[Valenti et al.(2008a)]{Valenti2008a}
Valenti, S., Benetti, S., Cappellaro, E., et al.\ 2008, \mnras, 383, 1485

\bibitem[Valenti et al.(2008b)]{Valenti2008b}
Valenti, S., Elias-Rosa, N., Taubenberger, S., et al.\ 2008, \apjl, 673, L155

\bibitem[Webbink(1984)]{Webbink1984}
Webbink, R.~F.\ 1984, \apj, 277, 355

\bibitem[Wheeler \& Hansen(1971)]{Wheeler1971}
Wheeler, J.~C., \& Hansen, C.~J.\ 1971, \apss, 11, 373

\bibitem[Whelan \& Iben(1973)]{Whelan1973}
Whelan, J., \& Iben, I., Jr.\ 1973, \apj, 186, 1007

\bibitem[Woosley \& Bloom(2006)]{Woosley2006}
Woosley, S.~E., \& Bloom, J.~S.\ 2006, \araa, 44, 507

\bibitem[Yaron \& Gal-Yam(2012)]{Yaron2012}
Yaron, O., \& Gal-Yam, A.\ 2012, \pasp, 124, 668

\end{thebibliography}
\end{document}